\documentclass[aps,prc,floatfix,showpacs,twocolumn, preprintnumbers]{revtex4-1}

\usepackage{caption}
\usepackage{subcaption}
\usepackage{graphicx}
\usepackage{amssymb}
\usepackage{amsmath}
\usepackage{cancel}
\usepackage[dvipsnames,usenames]{xcolor}
\usepackage{placeins}

\begin{document}
\preprint{FERMILAB-PUB-21-375-T}
\title{Electron scattering on ${\mathbf{A=3}}$ nuclei from quantum Monte Carlo based approaches}
\author{ {Lorenzo} Andreoli$^{\, {\rm a} }$,
{Joseph} Carlson$^{\, {\rm b} }$,
{Alessandro} Lovato$^{\, {\rm c,d,e} }$,
{Saori} Pastore$^{\, {\rm a} }$,
{Noemi} Rocco$^{\, {\rm f} }$,
{R. B.} Wiringa$^{\, {\rm c} }$
}
\affiliation{
$^{\,{\rm a}}$\mbox{Department of Physics and the McDonnell Center for the Space Sciences at Washington University in St. Louis, MO, 63130, USA}\\
$^{\,{\rm b}}$\mbox{Los Alamos National Laboratory, Los Alamos, NM, 87545 USA}\\
$^{\,{\rm c}}$\mbox{Physics Division, Argonne National Laboratory, Argonne, Illinois 60439, USA}\\
$^{\,{\rm d}}$\mbox{Computational Science Division Division, Argonne National Laboratory, Argonne, Illinois 60439, USA}\\
$^{\,{\rm e}}$\mbox{INFN-TIFPA Trento Institute of Fundamental Physics and Applications, Via Sommarive, 14, 38123 {Trento}, Italy}\\
$^{\,{\rm f}}$\mbox{Theoretical Physics Department, Fermi National Accelerator Laboratory, P.O. Box 500, Batavia, IL 60510, USA}\\
}
\date{\today}

\begin{abstract}
We perform first-principle calculations of electron-nucleus scattering on $^3$He and $^3$H using the Green's function Monte Carlo method and two approaches based on the factorization of the final hadronic state: the spectral-function formalism and the short-time approximation. These three methods are benchmarked among each other  and compared to the experimental data for  the longitudinal and transverse electromagnetic response functions of $^3$He, and the inclusive cross sections of both $^3$He and $^3$H. Since these three approaches are based on the same description of nuclear dynamics of the initial target state, comparing their results enables a precise quantification of the uncertainties inherent to factorization schemes. At sufficiently large values of the momentum transfer, we find an excellent agreement of the Green's function Monte Carlo calculation with experimental data and with both the spectral-function formalism and the short-time approximation. We also analyze the relevance of relativistic effects, whose inclusion becomes crucial to explain data at high momentum and energy transfer. 
\end{abstract}

\maketitle

\section{Introduction}
The spectrum of the inclusive lepton-nucleus cross section exhibits a variety of features that are sensible to both long- and short-range nuclear dynamics. At low energies, coherent scattering, excitation of low-lying nuclear states, and collective modes are the dominant reaction mechanisms. At energy transfers on the order of hundreds of MeV, the leading mechanism is quasielastic (QE) scattering, where the electroweak probe interacts primarily with individual bound nucleons. These, after interacting with other nucleons, are ejected from the target. Corrections to this leading one-body mechanism arise from processes in which the lepton couples to pairs of correlated  nucleons via nuclear  two-body currents~\cite{Carlson:1997qn,Benhar:2006wy,Bacca:2014tla}. 

Electron-scattering experiments play a key role in validating the nuclear shell model, as well as in exposing its limitations. A fully quantitative description of experimental data requires including nuclear correlations, which reduce the occupation probability of low-momentum shell-model states and lead to the appearance of high-momentum components in the nuclear wave function. These are generated by the short-range component of the nuclear interaction, also relevant for the stability of neutron stars~\cite{Benhar:2019rro,Piarulli:2019pfq,Sabatucci:2020xwt}. The experimental investigation of short-range correlations (SRC) has flourished over the last few years and it is realized by  selecting kinematics where the role of short-range correlated pairs of nucleons become dominant~\cite{Duer:2018sxh,Schmidt:2020kcl}. The analysis of experimental data taken in this kinematic region has unveiled the importance of the tensor component of the nuclear potential that causes the dominance of neutron-proton correlated pairs with respect to the proton-proton and neutron-neutron pairs~\cite{Wiringa:2013ala,Hen:2014nza,Atti:2015eda}. In addition, the analysis of SRC pairs is relevant to improve our  understanding of the interplay between nucleonic and partonic degrees of freedom~\cite{Hen:2016kwk,Schmookler:2019nvf}.

Moreover, current and planned neutrino-oscillation experiments rely on theoretical estimates of lepton-nucleus cross sections required to reconstruct the energy of the incoming neutrino~\cite{mb_web,nova_web,t2k_web,dune_web,hk_web}. The uncertainty associated with these calculations, often based on simplified models of nuclear dynamics, is
one of the most important sources of systematic error in these experiments. Therefore, achieving a robust description of all the reaction mechanisms at play in the broad kinematic region relevant to accelerator-based oscillation experiments is necessary to improve the accuracy of the extracted oscillation parameters~\cite{Benhar:2015wva,Katori:2016yel,Alvarez-Ruso:2017oui}.

Microscopic calculations of lepton-nucleus scattering cross sections use nucleons as fundamental degrees of freedom and 
allow one to fully account for the important many-body effects in both many-nucleon interactions and electroweak currents. 
A microscopic approach that has been extensively used in recent year  
is the quantum Monte Carlo (QMC) method~\cite{Carlson:2014vla}.  The QMC method, in particular
the Green's function Monte Carlo (GFMC), produces results for the inclusive cross sections of $^4$He and $^{12}$C
that are in excellent agreement with available experimental data~\cite{Lovato:2016gkq,Lovato:2017cux,Lovato:2020kba}. 
In a QMC calculation, nucleons are correlated in pairs and triplets via two- and three-nucleon interactions, while 
one- and two-nucleon electromagnetic currents are used to describe the interaction with the external probe. 
Two-body currents are constructed from the two-nucleon interactions consistently, that is by imposing that they 
satisfy charge conservation with the given two-nucleon interaction. The growing computational cost of the GFMC 
with the number of nucleons currently limits the applicability of this method to nuclei with $A \leq 12$ nucleons. Furthermore,
despite having relativistic corrections included in the electromagnetic current operator, the GFMC method is not 
applicable to lepton-nucleus scattering in the large momentum and energy transfer regime, where fully relativistic currents and kinematics must be considered. 

An alternative approach that addresses these shortcomings relies on the spectral function (SF) of the 
nucleus~\cite{Benhar:1994hw,Rocco:2015cil,Rocco:2018mwt}. This method is based on the factorization of the 
final hadronic states and has the advantage of being applicable to larger nuclear systems. Moreover, it can accommodate 
for both relativistic kinematics and meson-production mechanism~\cite{Rocco:2015cil,Rocco:2018mwt}. The latter has been 
included in the SF formalism by using the electroweak pion production amplitudes generated within the dynamical coupled-channel (DCC) model~\cite{Kamano:2013iva,Nakamura:2015rta,Kamano:2016bgm}. The contributions of these different reaction mechanisms have been 
combined to obtain neutrino and electron scattering on $^{12}$C for different kinematics ~\cite{Rocco:2019gfb, Rocco:2020jlx}.

Recently, the Short-time-approximation (STA)~\cite{Pastore:2019urn} has been developed to calculate nuclear responses 
in nuclei with $A>12$ within a QMC framework. At present, this computational algorithm has been tested within
the Variational Monte Carlo (VMC) method~\cite{Carlson:2014vla} to study electron scattering from the alpha particle.
The algorithm exploits a factorization scheme to consistently retain two-body physics, namely two-body currents and 
correlations. Despite limiting the description of the scattering process to interactions of the probe with pairs
of correlated nucleons, the STA is found to be in good agreement with both GFMC predictions and experimental data 
for electron scattering from the alpha particle~\cite{Pastore:2019urn} and, as in the SF formalism, can 
accommodate relativistic effects and meson production reactions. 

In this work, we compute the electromagnetic responses and inclusive double-differential electrons-scattering cross 
sections of $^3$H and $^3$He, comparing the GFMC, SF, and STA predictions with experimental data. Besides the 
intrinsic interest of our first-principle calculations, we gauge the accuracy and the regime of validity of 
the factorization approximation by comparing SF and STA results against {\it virtually exact} 
GFMC calculations that are carried out within the same model of nuclear dynamics for the initial state. In addition, we compare the 
GFMC and STA non-relativistic calculations with the relativistic results obtained within the SF formalism
(see, {\it e.g.}, Ref.~\cite{Rocco:2018tes}), and assess the importance of including relativistic effects at
kinematics regions with higher values of energy and momentum transfer. 

This paper is structured as follows. Sec.~\ref{sec:cross_section} provides the definition of the electromagnetic responses and inclusive cross section. Sec.~\ref{sec:theoretical} is devoted to the description of the GFMC, SF and STA approaches. Our results are summarized and discussed in Sec.~\ref{sec:results}, while in Sec.~\ref{sec:conclusions} we state our concluding remarks.

\section{Electron-nucleus scattering cross section}
\label{sec:cross_section}
The inclusive double differential cross section for the scattering of an electron with initial four-momentum $k=(E, {\bf k})$ on a nucleus at rest is written as 
\begin{align}
\Big(\frac{d^2\sigma}{d E^\prime d\Omega^\prime}\Big)_e & =\frac{\alpha^2}{Q^4}\frac{E^\prime}{E}L_{\mu\nu}R^{\mu\nu}\, ,
\label{xsec:em1}
\end{align}
where the outgoing electron has a momentum $k^\prime=(E^\prime, {\bf k}^\prime)$, $\alpha\simeq1/137$ is the fine structure constant, 
and $\Omega^\prime$ is the scattering solid angle in the direction specified by ${\bf k}^\prime$.
The energy and the momentum transfer are denoted by $\omega$  and {\bf q}, respectively, with $Q^2=-q^2={\bf q}^2-\omega^2$.
The lepton tensor is fully determined by the lepton kinematic variables. Neglecting the electron mass, it is given by 
\begin{equation}
L_{\mu \nu}  = \frac{1}{EE^\prime} (k_\mu k^\prime_\nu + k^\prime_\mu k_\nu - g_{\mu\nu}\, k \cdot k^\prime )\, .
\label{eq:lepton_def}
\end{equation}
The hadronic tensor describes the transition between the initial and final nuclear states $|\Psi_0\rangle$ and 
$|\Psi_f\rangle$ with energies $E^A_0$ and $E^A_f$, where $A$ denotes the number of nucleons in the nucleus
\begin{align}
R^{\mu\nu}({\bf q},\omega)&= \sum_f \langle \Psi_0|J^{\mu \, \dagger}({\bf q},\omega)|\Psi_f\rangle \langle \Psi_f| J^\nu({\bf q},\omega) |\Psi_0 \rangle \nonumber\\
&\times\delta (E^A_0+\omega -E^A_f)\, .
\label{eq:had_tens}
\end{align}
The sum included all the possible hadronic final states, both bound and in the continuum, and $J^\mu({\bf q},\omega)$ is the nuclear current operator. 
For inclusive processes, the cross section of Eq.~\eqref{xsec:em1} only depends on the longitudinal and transverse response functions,
$R_L({\bf q}, \omega) \equiv R^{00}({\bf q}, \omega)$ and $R_T({\bf q},\omega)\equiv [R^{xx}({\bf q}, \omega) + R^{yy}({\bf q}, \omega)] / 2$, respectively
\begin{align}
\Big(\frac{d^2\sigma}{d E^\prime d\Omega^\prime }\Big)_e &  =\left( \frac{d \sigma}{d\Omega^\prime} \right)_{\rm{M}} \Big[  \Big( \frac{q^2}{{\bf q}^2}\Big)^2  R_L(|{\bf q}|,\omega) \nonumber\\
&+ \Big(\tan^2\frac{\theta}{2}-\frac{1}{2}\frac{q^2}{{\bf q}^2}\Big)  R_T(|{\bf q}|,\omega) \Big] \ .
\label{eq:x:sec}
\end{align}
The Mott cross section 
\begin{align}
\label{Mott}
\left( \frac{d \sigma}{d \Omega^\prime} \right)_{\rm{M}}= \left[ \frac{\alpha \cos(\theta/2)}{2 E^\prime\sin^2(\theta/2) }\right]^2
\end{align} 
only depends upon the scattering angle $\theta$ and on the outgoing electron energy $E^\prime$.

\subsection{Nuclear current}

The GFMC, SF, and STA methods use as input a many-body nuclear Hamiltonian that consists of non-relativistic single-nucleon kinetic energy terms, 
and two- and three-nucleon interactions
\begin{equation}
H = \sum_i -\frac{\hbar^2}{2m} \,{\nabla}_i^2+ \sum_{i<j} v_{ij} + \sum_{i<j<k} V_{ijk} \ ,
\end{equation}
where $v_{ij}$ and $V_{ijk}$ are sophisticated potentials~\citep{Carlson:2014vla,Bacca:2014tla} that model the interaction between pairs and triples of nucleons. In this work, the Argonne $v_{18}$ two-nucleon interaction~\citep{Wiringa:1994wb} is utilized in combination with the Illinois-7 three-nucleon force~\citep{Pieper:2008rui}
(used in the GFMC calculations) or the Urbana IX three-nucleon interaction~\citep{Pudliner:1995wk} (used in the STA calculations). 
The highly-realistic Argonne $v_{18}$~\citep{Wiringa:1994wb} potential reflects the rich features of the nucleon-nucleon force. It is written in terms of operator structures involving space, momentum, spin and isospin nucleonic coordinates, predominantly arising from one- and two-meson-exchange-like mechanisms. The long-range part of the nucleon-nucleon interaction is due to one-pion-exchange; the intermediate-range component involves operator structures arising from multipion-exchange supported by phenomenological radial functions; the short-range part is described in terms of Woods-Saxon functions~\cite{Carlson:1997qn,Carlson:2014vla,Wiringa:1994wb}. The Argonne $v_{18}$ has 40 parameters that have been adjusted to fit the Nijmegen $pn$ and $pp$ scattering data base~\cite{Stoks:1993tb}, consisting of $\sim 4300$ data in the range of $0-350$ MeV, with a $\chi^2$/datum close to one. While fitting data up to 350 MeV, the Argonne $v_{18}$ reproduces the nucleon-nucleon phase shifts up to $\sim 1$ GeV, an indication that its regime of validity extends beyond the energy range utilized to constrain the adjustable parameters. This is also an indication that relativistic effects are (partially) embedded in the parameters entering the nucleon-nucleon interaction. 

Analogously to the nuclear potentials, electroweak currents can also be expressed as an expansion in many-body operators that act on nucleonic degrees of freedom
\begin{equation}
J^\mu=\sum_i j^\mu(i)+\sum_{i<j}j^\mu(ij) +\cdots \, .
\end{equation}
The ellipsis denotes terms involving three nucleons or more, which are found to be small~\cite{Marcucci:2005zc} and will be neglected in this work. 

The electromagnetic current can be schematically written as 
\begin{equation}
    J^\mu_{\rm EM}=J^\mu_{\gamma,S}+J^\mu_{\gamma,z}\, ,
\end{equation} 
where the first term is isoscalar and the second is isovector, depending upon the isospin operators $\tau_z$. The relativistic expression of the one-body current is
\begin{align}
j^\mu_{\rm EM}=\bar{u}({\bf p}^\prime)\Big[ {\mathcal F}_1 \gamma^\mu+ i \sigma^{\mu\nu}q_\nu \frac{{\mathcal F}_2}{2m_N}\Big] u({\bf p})\, ,
\label{rel:1b:curr}
\end{align}
where ${\bf p}$ and ${\bf p}^\prime$ are the initial and final nucleon momentum. The isoscalar (S) and isovector (V) form factors, ${\mathcal F}_1$ and ${\mathcal F}_2$,
are given by combinations of the Dirac and Pauli ones, $F_{1}$ and $F_2$, as
\begin{align}
{\mathcal F}_{1,2}= \frac{1}{2}[F_{1,2}^S+F_{1,2}^V\tau_z]\, ,
\end{align}
where $\tau_z$ is the isospin operator and 
\begin{align}
F_{1,2}^S=F_{1,2}^p + F_{1,2}^n\, , \ \ \ \ \ \ \ F_{1,2}^V=F_{1,2}^p - F_{1,2}^n\, . 
\label{f12:iso}
\end{align}
The Dirac and Pauli form factors can be expressed in terms of the electric and magnetic form factors of the proton and neutron as
\begin{align}
F_1^{p,n}=&\frac{G_E^{p,n}+\tau G_M^{p,n}}{1+\tau}\, ,\ \ \ \ \ \ \ \ F_2^{p,n}=\frac{G_M^{p,n}-G_E^{p,n}}{1+\tau}\, ,
\end{align}
with $\tau=Q^2/4m_N^2$. 
While the SF formalism uses the relativistic expressions above, the one-body charge and current operators employed in the GFMC and STA approaches are obtained from the nonrelativistic reduction of the covariant operator of Eq.~\eqref{rel:1b:curr}, including all the terms up to $1/m_N^2$ in the expansion. The charge (0), transverse ($\perp$), and longitudinal ($\parallel$) components with respect to the three-momentum ${\bf q}$ of the nonrelativistic expansion of the current read 
\begin{align}
j^0_{\gamma,S}(i)=& \frac{G_E^S}{2\sqrt{1+Q^2/4m_N^2}}-i\frac{2 G_M^S-G_E^S}{8m^2_N}{\bf q}\cdot (\boldsymbol{\sigma}_i\times {\bf p}_i)\, , \nonumber\\
{\bf j}^\perp_{\gamma,S}(i)=&\frac{G_E^S}{2m_N}{\bf p}^\perp_i-i\frac{G_M^S}{4m_N}({\bf q}\times\boldsymbol{\sigma})_i\, , \nonumber\\
j^\parallel_{\gamma,S}(i)=&\frac{\omega}{|{\bf q}|}j^0_{\gamma,S}\, .
\label{isosc:curr}
\end{align}
The isoscalar and isovector component of the electric and magnetic form factors are
\begin{align}
G_{E,M}^S &=G_{E,M}^p+G_{E,M}^n\nonumber\\ G_{E,M}^V& =G_{E,M}^p-G_{E,M}^n\, .
\end{align}
The isovector contributions to the current $J^\mu_{\gamma,z}$ are obtained by replacing $G_{E,M}^S\rightarrow G_{E,M}^V \tau_z$. 

The gauge invariance of the theory imposes that the electromagnetic charge and current operators satisfy the continuity equation
\begin{equation}
\mathbf{q}\cdot \mathbf{J}_\text{EM}=[ H, \rho_\text{EM}]\, ,
\end{equation}
where $\rho_\text{EM}\equiv J^0_\text{EM}$, which provides an explicit
connection between the nuclear interactions and the longitudinal
component of the current operators. For instance, the isospin and
momentum dependence of the NN interactions leads to nonvanishing
commutators with the one-body charge operator and hence to the emergence
of two-body terms in the current operator. The GFMC and STA calculations reported 
in this work have been carried out using the two-body currents most recently summarized in Refs.~\cite{Bacca:2014tla,Carlson:1997qn,Carlson:2014vla}. 
They include both ``model-independent'' and ``model-dependent'' terms as defined in Ref.~\cite{Riska:1989bh}.
The former are obtained from the nucleon-nucleon interaction, and satisfy current conservation by construction. 
The leading operator is the isovector ``$\pi$-like'' current, with important contributions also due to $\rho$-like terms. 
The additional two-body currents induced by the momentum dependence of the nucleon-nucleon interaction have been found 
to give contributions that are much smaller than those generated by the static part of this interaction, in particular 
the OPE current~\cite{Marcucci:2008mg}.

The transverse components of the two-body currents, {\it i.e.}, the model-dependent components,  cannot be directly linked to the nuclear Hamiltonian. 
In this work, we adopt the latest formulation of Refs.~\cite{Shen:2012xz,Lovato:2013cua,Lovato:2015qka,Lovato:2016gkq,Lovato:2017cux,Pastore:2019urn,Lovato:2020kba} 
and include the isoscalar $\rho\pi\gamma$ transition and the isovector current associated with the excitation of intermediate $\Delta$-isobar resonances. 
The $\rho\pi\gamma$  couplings are extracted from the widths of the radiative decay $\rho\to \pi\gamma$~\cite{Berg:1980lwp} and the $Q^2$ 
dependence of the electromagnetic transition form factor is modeled assuming vector-meson dominance. Among the model-dependent currents, 
those associated with the $\Delta$ isobar are the most important and enhance the transverse electromagnetic response functions. 

It is worthwhile to point out that the realistic interactions and currents utilized in the present work---Argonne $v_{18}$ two-nucleon~\cite{Wiringa:1994wb} and Urbana IX or Illinois-7 three-nucleon~\cite{Pudliner:1995wk,Pieper:2008rui} interactions and associated currents---provide a quantitatively successful description of many nuclear electroweak
observables~\cite{Bacca:2014tla}, including charge radii, electromagnetic moments and transition rates, charge and magnetic form factors of nuclei with up to $A=12$ nucleons~\cite{Carlson:1997qn,Marcucci:2005zc,Marcucci:2008mg,Lovato:2013cua,Pastore:2009is,Girlanda:2010vm,Pastore:2011ip,Pastore:2012rp,Datar:2013pbd,Pastore:2014oda,Pastore:2017uwc},
and electromagnetic response functions~\cite{Shen:2012xz,Lovato:2015qka,Lovato:2016gkq,Lovato:2017cux,Pastore:2019urn,Lovato:2020kba,Schiavilla:2018udt,Barrow:2020mfy}.

\section{Theoretical approaches}
\label{sec:theoretical}

\subsection{Green's function Monte Carlo}
The Green's function Monte Carlo method is suitable to solve the Schr\"odinger equation of nuclei with up to $A=12$ nucleons with percent-level accuracy. The ground-state of a given Hamiltonian $H$ is obtained by propagating in imaginary-time a starting trial wave function $|\Psi_T\rangle$
\begin{equation}
|\Psi_0\rangle \propto\lim_{\tau\to\infty}\exp[-(H-E_0)\tau]|\Psi_T\rangle\,,
\end{equation}
where $\tau$ is the imaginary time, and $E_0$ is a parameter used to control the normalization. The above imaginary-time propagation can also be used to extract dynamical properties of atomic nuclei. The energy dependence of the response functions can be inferred by computing their Laplace transform, dubbed as Euclidean response function~\cite{Carlson:2001mp}
\begin{equation}
E_\alpha(\mathbf{q},\tau) = \int_{\omega_{\rm th}}^\infty d\omega\, e^{-\omega \tau} R_\alpha (\mathbf{q},\omega), \quad \alpha=L,T\, .
\end{equation}
Fixing the intrinsic energy dependence of the charge and current operators to the QE peak: $J_\alpha ({\bf q}) \equiv J_\alpha({\bf q},\omega_{\rm QE})$, with $\omega_{\rm QE}=\sqrt{{\bf q}^2+m_N^2}-m_N$, one can express the Euclidean responses as ground-state expectation values
\begin{align}
E_\alpha({\bf q},\tau) &= \langle \Psi_0 | J_\alpha^\dagger ({\bf q}) e^{-(H-E_0)\tau} J_\alpha({\bf q}) | \Psi_0\rangle \nonumber\\
&- |F_\alpha(\mathbf{q})|^2 e^{-\omega_{el}\tau}\, ,
\label{eq:eucl_elastic}
\end{align}
where the elastic form factor is defined as $F_\alpha(\mathbf{q}) = \langle \Psi_0 | J_\alpha({\bf q})| \Psi_0 \rangle$. The calculation of the imaginary-time correlation operator $\langle \Psi_0 | J_\alpha^\dagger ({\bf q}) e^{-(H-E_0)\tau} J_\alpha({\bf q}) | \Psi_0\rangle$ is carried out with GFMC methods similar to those used in projecting out the exact ground state of $H$ from a trial
wave function. It proceeds in two steps. First, an unconstrained imaginary-time propagation of the state $|\Psi_0\rangle$ is performed and stored. Then, the states $J_\alpha({\bf q})|0\rangle$ are evolved in imaginary time following the path previously saved. During this latter imaginary-time evolution, estimates for $E_\alpha({\bf q},\tau_i)$ on a uniform grid of $\tau_i$ values are obtained by evaluating the scalar products of $e^{-(H-E_0)\tau_i} J_\alpha({\bf q})|0\rangle$ with $\langle \Psi_0 |J_\alpha^\dagger ({\bf q})$ --- a complete discussion of the methods is in Refs.~\cite{Lovato:2015qka,Lovato:2016gkq,Lovato:2017cux}.

The above expectation value is evaluated on a uniform grid of $n_\tau$ imaginary-time points~\cite{Carlson:1992ga,Carlson:2001mp}. A set of noisy estimates for $E_\alpha({\bf q},\tau_i)$ can be obtained by performing independent imaginary-time propagations, from which the average Euclidean response $\bar{E}_\alpha({\bf q},\tau_i)$ and the covariance $C_{ij}$ between the data at $\tau=\tau_i$ and $\tau=\tau_j$ can be readily estimated~\cite{Lovato:2016gkq}. Note that, in general, the covariance matrix $C$ is nondiagonal because of correlations among the imaginary-time points.

Retrieving the energy dependence of the response functions from their Euclidean counterparts is a nontrivial problem. For the smooth quasiealstic responses on which this work focuses on, we employ a version of the maximum-entropy technique developed specifically for this type of problem~\cite{Lovato:2015qka}. It has to be noted that machine-learning algorithms have recently been developed to invert the Laplace transform~\cite{Raghavan:2020bze} and are capable of precisely reconstructing the low-energy transfer region of the response functions. 

\subsection{Short-time approximation}

In the short-time approximation~\cite{Pastore:2019urn}, the response defined in Eq.~(\ref{eq:had_tens}) is calculated performing 
a real-time propagation. This scheme can be appreciated by rewriting the response as
\begin{eqnarray}
 R_\alpha ({\bf q},\omega) &=&
  \int_{-\infty}^\infty  \frac{d t}{2 \pi} \,
  {\rm e}^{ i \left(\omega+E_0\right)   t }\, \nonumber \\
  &\times&\,\langle \Psi_0 |
 	  J_\alpha^\dagger ({\bf q})\,{\rm e}^{-i H t}\,
 	  J_\alpha ({\bf q}) |\Psi_0 \rangle \ ,
\label{eq:realtime}
\end{eqnarray}
where we have replaced the sum over the final states with a real-time propagator.
In the STA, we evaluate the real-time matrix element in Eq.~(\ref{eq:realtime}) 
for short times. We retain the full QMC ground state and current operators,
and final state interactions at the two-nucleon level---specifically, those final state
interactions affecting only pairs involved at the electromagnetic interaction vertex. 
In doing so, the STA accounts for two-nucleon interactions and currents and ensuing 
interference terms, consistently, {\it i.e.}, satisfying current conservation.
In practice, in the scattering process only two correlated
nucleons interact with the probe via both one- and two-body currents. 
Schematically, the current-current correlator entering the 
real-time matrix elements can be written as
\begin{eqnarray}
&& J^\dagger\, {\rm e}^{-iHt}\,  J 
=\sum_i J_i^\dagger\,  {\rm e}^{-iHt} J_i 
	+ \sum_{i \neq j} J_i^\dagger\,  {\rm e}^{-iHt} J_j \\
	&&+ \sum_{i\neq j}
	\left( J^\dagger_{i}{\rm e}^{-iHt} J_{ij} + J^\dagger_{ij}{\rm e}^{-iHt} J_{i}   + J^\dagger_{ij} {\rm e}^{-iHt} J_{ij}\right)+\cdots \ , \nonumber
\label{eq:e25}
\end{eqnarray}
where we neglected terms terms with three or more active nucleons. 
In the expansion above, the Hamiltonian at the vertex correlates nuclei in pairs, that 
is it only includes the two-nucleon interaction which in this work is the Argonne $v_{18}$. 
Three-nucleon interaction effects are ignored in the 
final states, although they are fully included in the ground state. 
Note that only two nucleons are propagated. The equations 
above show how interference terms between one- and two-nucleon currents
are taken into account. While correctly reproducing the sum rules, the STA does not 
reproduce the correct threshold behavior of the response at values of momentum
transfer $q\lesssim 300$ MeV/c. We account for threshold effects by redistributing the 
strength of the response to higher values of energy transfer and preserving the 
correct value for the sum rules. Details on how threshold effects are accounted
for can be found in Ref.~\citep{Pastore:2019urn}. 

After the insertion of a complete set of two nucleon states, 
the response in~Eq.(\ref{eq:realtime}) can be evaluated as an integral of a 
response density, $D_\alpha(e,E_{\rm cm})$, over the relative energy, $e$, and center of mass energy, 
$E_{\mathrm cm}$, of the interacting pair (or, equivalently, over relative and CM momenta, ${\bf p}^\prime$ and ${\bf P}^\prime$, of the pair of struck nucleons):
\begin{eqnarray}
 R_\alpha ({\bf q},\omega) &=&
  \int_{0}^{\infty} d e \int_{0}^{\infty} d E_{\mathrm{cm}} \,
  D_\alpha\left(e, E_{\mathrm{cm}}\right)\, \nonumber \\
  &\times&\,\delta\left(\omega+E-e-E_{\mathrm{cm}}\right) \ .
\label{eq:density}
\end{eqnarray}
This response has contributions coming from the ground state for which 
the exact elastic response is simply $\propto |\langle \Psi_0 | J_\alpha({\bf q})| \Psi_0 \rangle|^2$.

\begin{figure}[h]
    \includegraphics[width=\columnwidth]{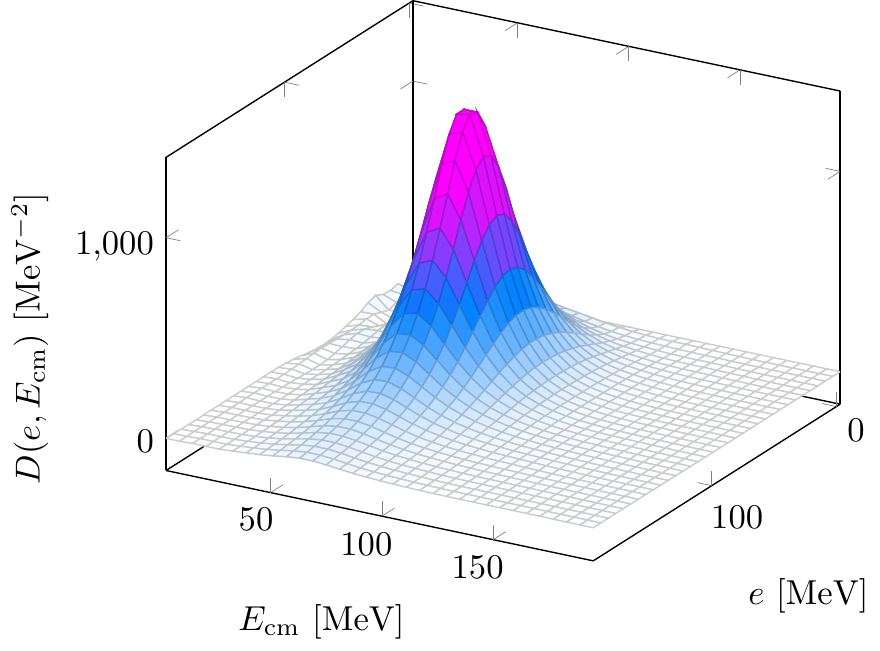} 
\caption{$^3$H total longitudinal response density at {q = 300 MeV/c}, as a function of relative energy and center-of mass energy.}
\label{fig:density_total}
\end{figure}

\begin{figure}[h]
    \includegraphics[width=\columnwidth]{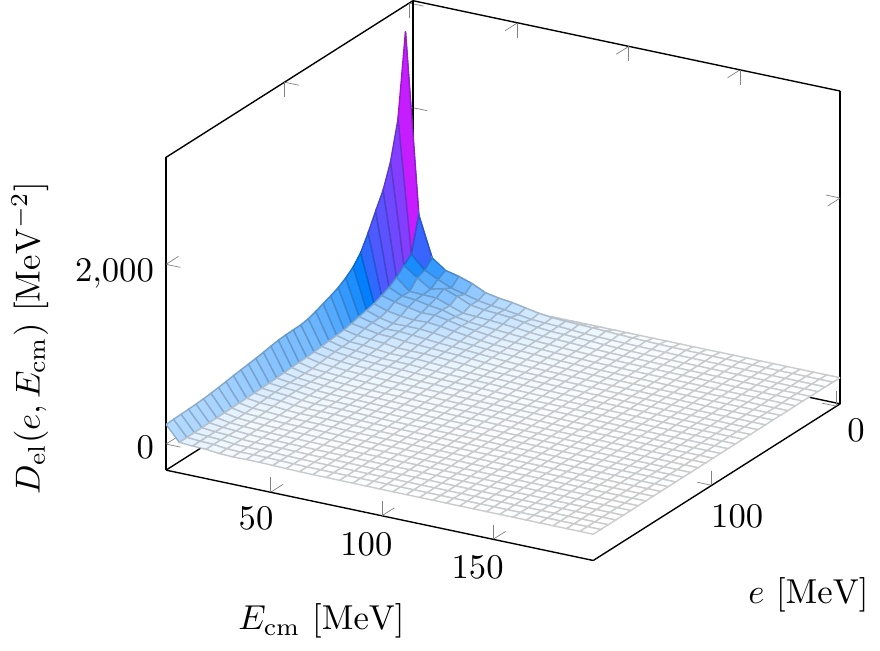} 
\caption{$^3$H elastic contribution to the total longitudinal response density at {q = 300 MeV/c.}}
\label{fig:density_elastic}
\end{figure}

Analogously to the subtraction of the elastic form factor in Eq.~(\ref{eq:eucl_elastic}),
the contribution of the exact elastic response---illustrated in Fig.~\ref{fig:density_elastic}---is subtracted
from the total response density---given in Fig.~\ref{fig:density_total}---that is 
\begin{equation}
D(e, E_{\mathrm{cm}}) - D_{\mathrm{el}}(e, E_{\mathrm{cm}}) \ ,
\end{equation}
where, for convenience, we omitted the subscript $\alpha$.
The elastic contribution $D_{\mathrm{el}}$ is calculated from the overlaps between the ground state and intermediate states with two active nucleons, here schematically denoted with $\Psi_2$.
Assuming that the elastic contribution is large at small momentum transfer and that that the internal nuclear dynamics dominates the overlap to states of given ${\bf p}^\prime$ and ${\bf P}^\prime$,  the elastic contribution to the total response density can be approximated with
\begin{align}
&D_{\mathrm{el}}({\bf q}, {\bf p}^\prime, {\bf P}^\prime) = \,
 \left|\left\langle \Psi_0 | J \left(\mathbf{q}\right) \mid \Psi_0 \right\rangle\right|^{2}\nonumber \\
&\times \sum_{\beta} \langle \Psi_0| \Psi_{2}\left({\bf p}^\prime, {\bf P}^\prime, \beta \right)\rangle \langle \Psi_{2}\left({\bf p}^\prime, {\bf P}^\prime, \beta \right) |\Psi_0 \rangle \ ,
\end{align}
where the sum runs over all two body quantum numbers $\beta$.
In the limit of zero momentum transfer the response is fully elastic, while at large momentum transfer the elastic response goes to zero.

\subsection{Spectral function}

In the region of large ${\bf q}$ it is reasonable to approximate the hadronic final state with the factorized expression
\begin{equation}
    |\Psi_f\rangle = |{\bf p}\rangle \otimes |\Psi_n^{A-1}\rangle\, ,
\end{equation}
where $|p\rangle$ is a plane wave describing the propagation of the final state nucleon with momentum $|{\bf p}|$, while $|\Psi_n^{A-1}\rangle$ describes the $(A-1)$-body spectator system. 
The incoherent contribution to the longitudinal and transverse response function is obtained by inserting a single-nucleon completeness relation in Eq.~\eqref{eq:had_tens}
\begin{align}
   & R_\alpha({\bf q}, \omega)= \sum_{\tau_k=p,n} \int  \frac{d^3k}{(2\pi)^3} dE \Big[ P_{\tau_k}({\bf k},E) \nonumber\\
    &\qquad  \times \frac{m_N^2}{e({\bf k})e({\bf k+q})} \sum_i \langle k| j^\dagger_{i,\alpha}|k+q \rangle \langle p| j_{i,\alpha}| k \rangle \nonumber\\
    &\qquad  \times \delta(\tilde{\omega}+e({\bf k})-e({\bf k+q}))\Big]
\end{align}
where we retain only the one-body current contribution. In the above equation we used the relations ${\bf p}={\bf k+q}$ and $\tilde{\omega}=\omega -E +m_N- e({\bf k})$. 

The spectral function $P_{\tau_k}({\bf k},E)$, i.e., the probability distribution of removing a nucleon with momentum ${\bf k}$ and isospin $\tau_k=p,n$ from 
the target nucleus, leaving the residual $(A-1)$ system with an excitation energy $E$, can be written as~\cite{Benhar:1993ja}
\begin{align}
    P_{\tau_k}(\mathbf{k},E)&=\sum_n |\langle \Psi_0^A| [|k\rangle\, |\Psi_n^{A-1}\rangle]|^2 \nonumber\\
    & \times \delta(E+E^A_0-E_{n}^{A-1})\, .
\label{pke:hole}
\end{align}
Here $|\Psi_0^A\rangle$ is the ground state of the Hamiltonian, such that $H|\Psi_0^A\rangle = E_0 |\Psi_0^A\rangle$,
whereas $|\Psi_n^{A-1}\rangle$ are the eigenstates and energies of the $(A-1)$-nucleon system: 
$H|\Psi_{n}^{A-1}\rangle = E_n^{A-1} |\Psi_n^{A-1}\rangle$. 

Note that in Eq.~\eqref{pke:hole} the state $|k\rangle$ represents a single-particle plane wave with momentum $\mathbf{k}$ and isospin $\tau_k$ (we average over the spin). Using second-quantization definitions, we introduce the annihilation and creation operators $a_k, a^\dagger_k$, so that the definition of the SF reads
\begin{align}
    P_{\tau_k}(\mathbf{k},E)&=\sum_n |\langle \Psi_0^A| a_k^\dagger |\Psi_n^{A-1}\rangle|^2 \nonumber\\
    &\times \delta(E+E^A_0-E_{n}^{A-1})\, .
\end{align}
The single-nucleon momentum distribution is obtained integrating the spectral function over the removal energy
\begin{equation}
    n_{\tau_k}(\mathbf{k})=\langle\Psi_0^A|a^\dagger_k a_k|\Psi_0^A\rangle= \int dE P_{\tau_k}(\mathbf{k},E)\, .
\label{eq:nk_def}
\end{equation}
The proton and neutron spectral function and the corresponding momentum distributions are normalized as
\begin{align}
  \int dE \frac{d^3 k} {(2\pi)^3} P_p(\mathbf{k},E) = \int \frac{d^3 k} {(2\pi)^3} n_p(\mathbf{k}) &=  Z\, , \nonumber\\
    \int dE \frac{d^3 k} {(2\pi)^3} P_n(\mathbf{k},E) = \int \frac{d^3 k} {(2\pi)^3} n_n(\mathbf{k}) &=  A-Z\, ,  
\end{align}
where $Z$ is the number of protons and $A$ is the number of nucleons of a given nucleus. This normalization is consistent with
that of the VMC single-nucleon momentum distributions reported in~\cite{nk_web}.

For clarity, let us deal with the proton spectral function first. The single-nucleon (mean-field) 
contribution $P_p^{\rm MF}(\mathbf{k},E)$ for $A=3$ nuclei corresponds to identifying $|\Psi_n^{A-1}\rangle$ with the ground-state of the $A-1$ system
\begin{align}
& P_p^{\rm MF} (\mathbf{k},E)= \nonumber\\
&\qquad n_p^{\rm MF} (\mathbf{k})  \delta\Big(E-B_{A}+B_{A-1}-\frac{k^2}{2m_{A-1}}\Big)\, ,
\end{align}
where $B_{A}$ and $B_{A-1}$ are the binding energies of the initial and the $A-1$, remnant nucleus respectively and $m_{A-1}$ is the mass of the remnant. In the above equation we introduced the mean-field proton momentum distribution 
\begin{equation}
n_p^{\rm MF} (\mathbf{k})=| \langle \Psi_0^{A}| [|k\rangle \otimes |\Psi_n^{A-1}\rangle]|^2\,,
\end{equation}
in which $\langle \Psi_0^{A} | [|k\rangle \otimes |\Psi_n^{A-1}\rangle$ is the Fourier transform of the single-nucleon radial overlap that can be computed within both VMC and GFMC~\cite{overlaps_web}.

The high removal energy and momentum component of the spectral function arises from the contribution of correlated pairs of nucleons, as argued in Ref.~\cite{Benhar:1993ja} and,
more recently, in the context of the contact formalism~\cite{Weiss:2018tbu,Cruz-Torres:2019fum}. It amounts to a three-body final state with a high-momentum nucleon and a leftover pair of nucleons: 
$|\Psi_n^{A-1}\rangle \to |k^\prime\rangle \, |\Psi_n^{A-2}\rangle$ with $H |\Psi_n^{A-2}\rangle=E_{n}^{A-2}|\Psi_n^{A-2}\rangle$.
Note that we have neglected the correlations between the struck nucleon $|k^\prime\rangle$ and the pair of spectator nucleons. As a consequence, the state 
$|k^\prime\rangle \, |\Psi_n^{A-2}\rangle$ is not orthogonal to the ground state of $|\Psi_n^{A-1}\rangle$, entering the mean-field piece of the spectral function.

The corresponding two-body (correlation) contribution to the SF is given by
\begin{align}
P_p^{\rm corr} (\mathbf{k},E)&=\sum_n \int \frac{d^3 k^\prime}{(2\pi)^3} |\langle \Psi_0^A| [|k\rangle\,  |k^\prime\rangle \, |\Psi_n^{A-2}\rangle]|^2\nonumber\\
&\times \delta(E+E^A_0-e(\mathbf{k}^\prime)-E_{n}^{A-2})\, .
\label{pke:2b}
\end{align}
Assuming that the $(A -2)$-nucleon binding energy is narrowly distributed around a central value $\bar{B}_{A-2}$, we can use the completeness of the
final states $|\Psi_n^{A-2}\rangle$ to get
\begin{align}
& P_p^{\rm corr} (\mathbf{k},E)=  \mathcal{N}_p \sum_{\tau_{k^\prime}=p,n} \int \frac{d^3 k^\prime}{(2\pi)^3} \Big[ n_{p,\tau_{k^\prime}}(\mathbf{k},\mathbf{k}^\prime)\nonumber \\
&\qquad \times \delta \Big(E-B_A-e(\mathbf{k}^\prime)+\bar{B}_{A-2}-
\frac{(\mathbf{k}+\mathbf{k}^\prime)^2}{2m_{A-2}}\Big)\Big]\, , 
\label{pke:2b2}
\end{align}
where $m_{A-2}$ is the mass of the recoiling $A-2$ system and $\mathcal{N}_p$ is an appropriate normalization factor. VMC estimates of the two-nucleon momentum distribution, defined as
\begin{equation}
n_{\tau_k,\tau_{k^\prime}}(\mathbf{k},\mathbf{k}^\prime) = \langle \Psi_0^A| a^\dagger_k a_k a^\dagger_{k^\prime} a_{k^\prime} |\Psi_0^A\rangle\, ,
\end{equation}
can be found online~\cite{nkk_web} for several nuclei with up to $A=12$ nucleons. To isolate the contribution of short-range correlated nucleons, we introduce cuts on the relative distance between the pairs of particles in the two-body momentum distribution. The cut is selected in such a way that the overall normalization of the nucleon spectral function is correctly reproduced. Selecting pairs with a given range of relative distance provides an effective way to orthogonalize 
$|k^\prime\rangle \, |\Psi_n^{A-2}\rangle$ and the ground state of $|\Psi_n^{A-1}\rangle$ by isolating the contribution of short-range correlated nucleons in the ${A-2}$ system. 

\begin{figure}[t]
    \includegraphics[width=\columnwidth]{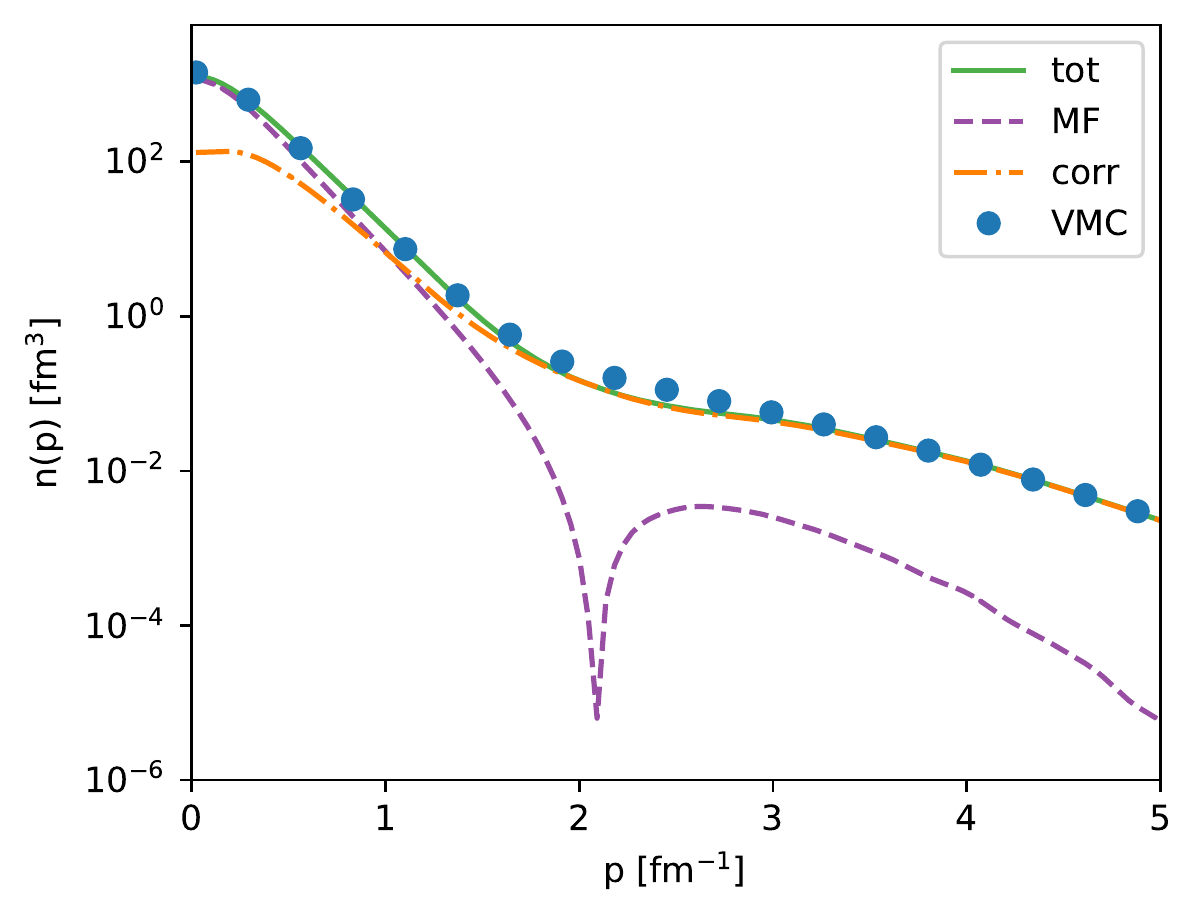} 
\caption{Single proton momentum distribution of $^3$He.}
\label{fig:nk_he3}
\end{figure}

The full SF is given by the sum of the mean-field and the correlated part~\cite{Benhar:1993ja}
\begin{equation}
P_p(\mathbf{k},E)= P_p^{\rm MF} (\mathbf{k},E) + P_p^{\rm corr} (\mathbf{k},E)\, .
\end{equation}
The momentum distributions obtained integrating the total, MF, and correlated spectral function of $^3$He are displayed in Fig.~\ref{fig:nk_he3} and compared to the VMC momentum distribution computed independently. As expected, the MF component of $n(k)$ dominates the low-momentum region, whereas the short-range correlated pairs entering $P_p^{\rm corr}$ mostly contribute to the high-momentum tails. The sum of the MF and correlation components of the momentum distribution is in excellent agreement with the VMC momentum distribution, corroborating the accuracy of our approach.

\subsection{Cross section calculation}

The analysis of scaling properties of nuclear response functions has been discussed in a number of works \cite{Alberico:1988bv,Barbaro:1998gu}. More recently the scaling ansatz has been successfully used to interpolate and extrapolate electroweak response functions for different values of the energy and momentum transfer~\cite{Rocco:2018tes,Lovato:2020kba,Barrow:2020mfy}. 
In the present work, we adopt the same interpolation algorithm based on the scaling of the nuclear responses presented in Ref.~\cite{Rocco:2018tes}. For the GFMC and STA calculations which rely on a nonrelativistic treatment of the kinematics, the longitudinal and transverse response functions can be expressed as a function of the nonrelativistic scaling variable $\psi_{\rm nr}$ defined as
\begin{equation}
    \psi_{\rm nr}= \frac{m_N}{|{\bf q}|k_F}\Big(\omega-\frac{|{\bf q}|^2}{2m_N}-\epsilon\Big)
\end{equation}
where $k_F$ is the Fermi Momentum and $\epsilon$ is a parameter chosen to account for binding effects in the initial and final states. In the present analysis we used $k_F=180$ MeV and $\epsilon= 10\ (15)$ MeV for the $^3$H ($^3$He) nucleus. \\
The SF formalism allows one to directly compute the cross section; the computational cost required to evaluate the nuclear response functions for different values of $\omega$ and ${\bf q}$ is quite small. In order to test the accuracy of the interpolation procedure, within the SF approach we compared the cross-section results obtained from a direct calculation and interpolating the response functions. Since the SF  quasielastic kinematics is fully relativistic, in this case we use the scaling function defined in Ref.~\cite{Caballero:2005sj}
\begin{equation}
    \psi=\frac{1}{\sqrt{\epsilon_F}}\frac{\lambda-\tau}{\sqrt{(1+\lambda)\tau+\kappa\sqrt{\tau(1+\tau)}}}
\end{equation}
with
\begin{align*}
    \epsilon_F=&\sqrt{k_F^2+m_N^2}/m_N-1\, , \nonumber\\
    \lambda=& (\omega-\epsilon)/(2 m_N)\, , \nonumber\\
    \kappa=& |{\bf q}|/(2 m_N)\, , \nonumber\\
    \tau=& (|{\bf q}|^2-\omega^2)/(4 m_N^2)\, .
\label{eq:epsf}
\end{align*}
The values of $k_F$ and $\epsilon$ are the sames as in the nonrelativistic case.

\section{Results}
\label{sec:results}
\begin{figure*}[h!]
    \begin{subfigure}[t]{0.48\textwidth}
    \includegraphics[width=\linewidth]{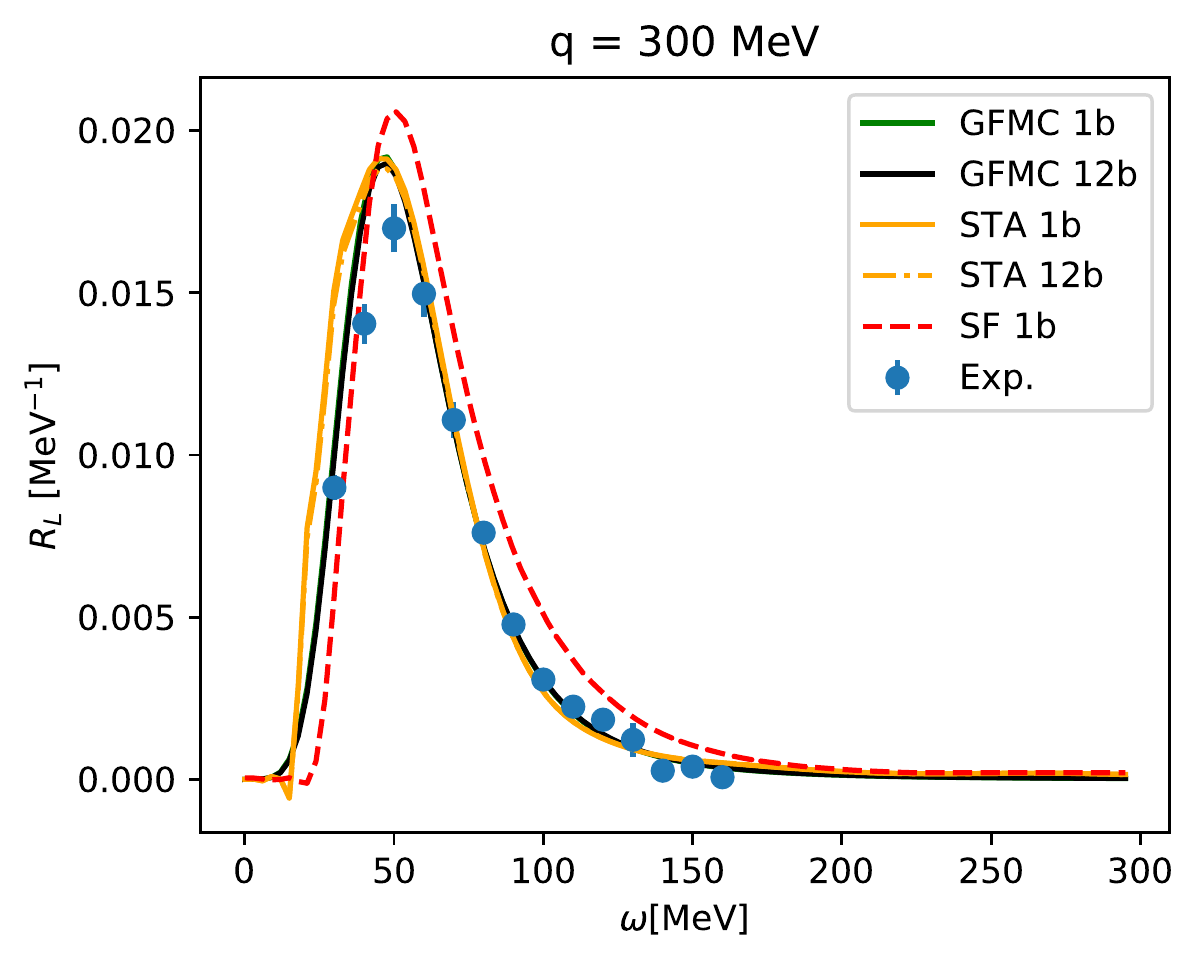} 
    \end{subfigure}
    \hspace{0.1cm}
    \begin{subfigure}[t]{0.48\textwidth}
    \includegraphics[width=\linewidth]{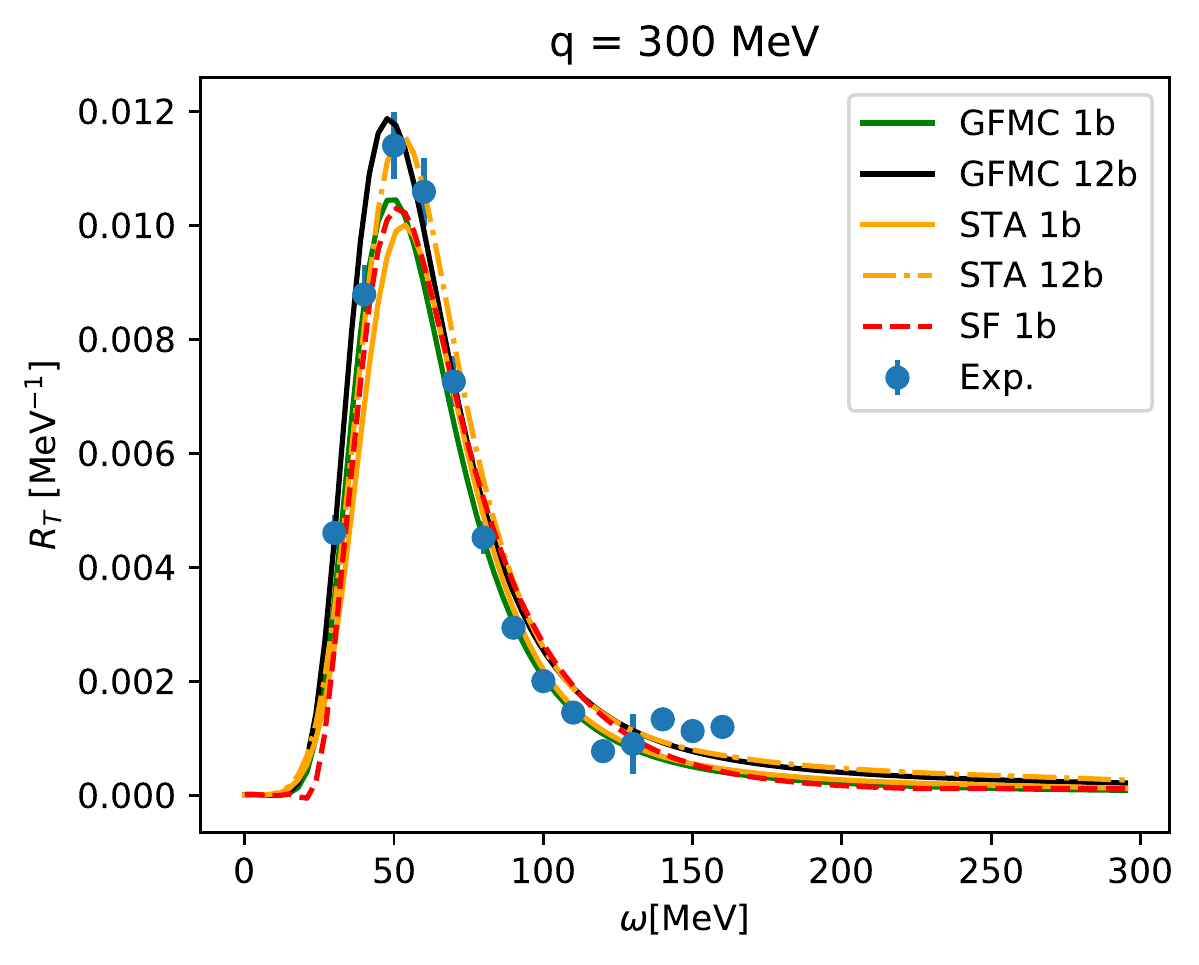} 
    \end{subfigure}
    \begin{subfigure}[t]{0.48\textwidth}
    \includegraphics[width=\linewidth]{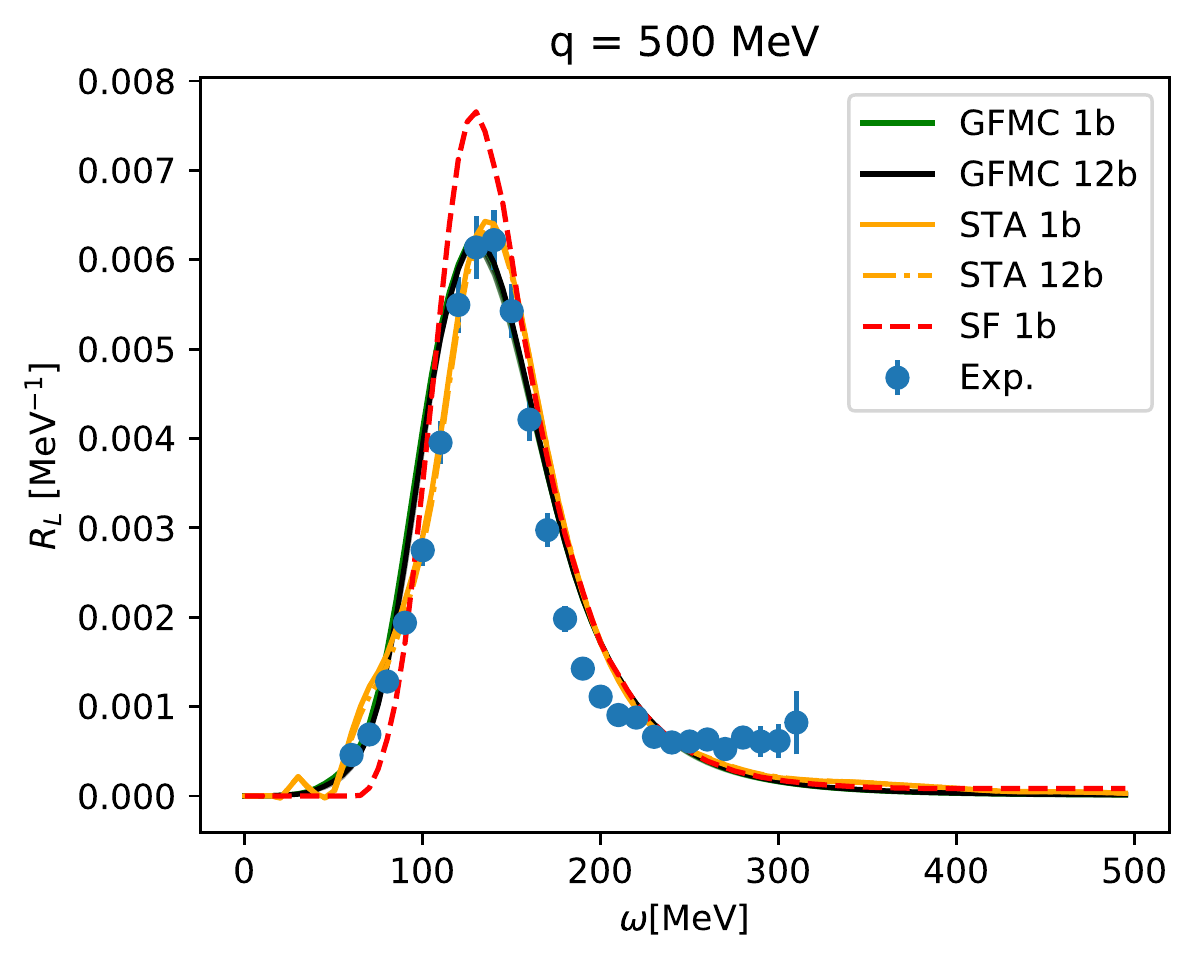} 
    \end{subfigure}
    \hspace{0.1cm}
    \begin{subfigure}[t]{0.48\textwidth}
    \includegraphics[width=\linewidth]{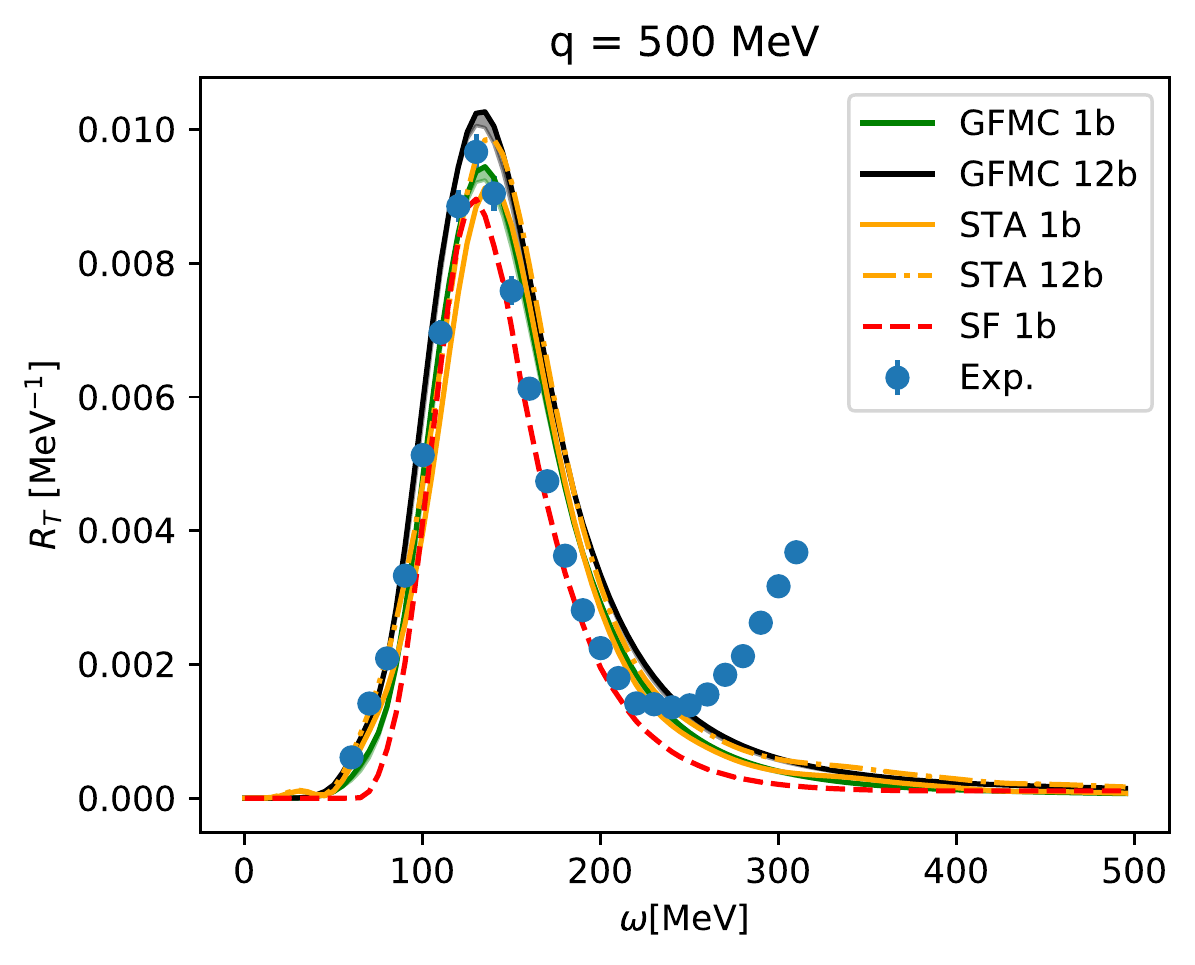} 
    \end{subfigure}
    \begin{subfigure}[t]{0.48\textwidth}
    \includegraphics[width=\linewidth]{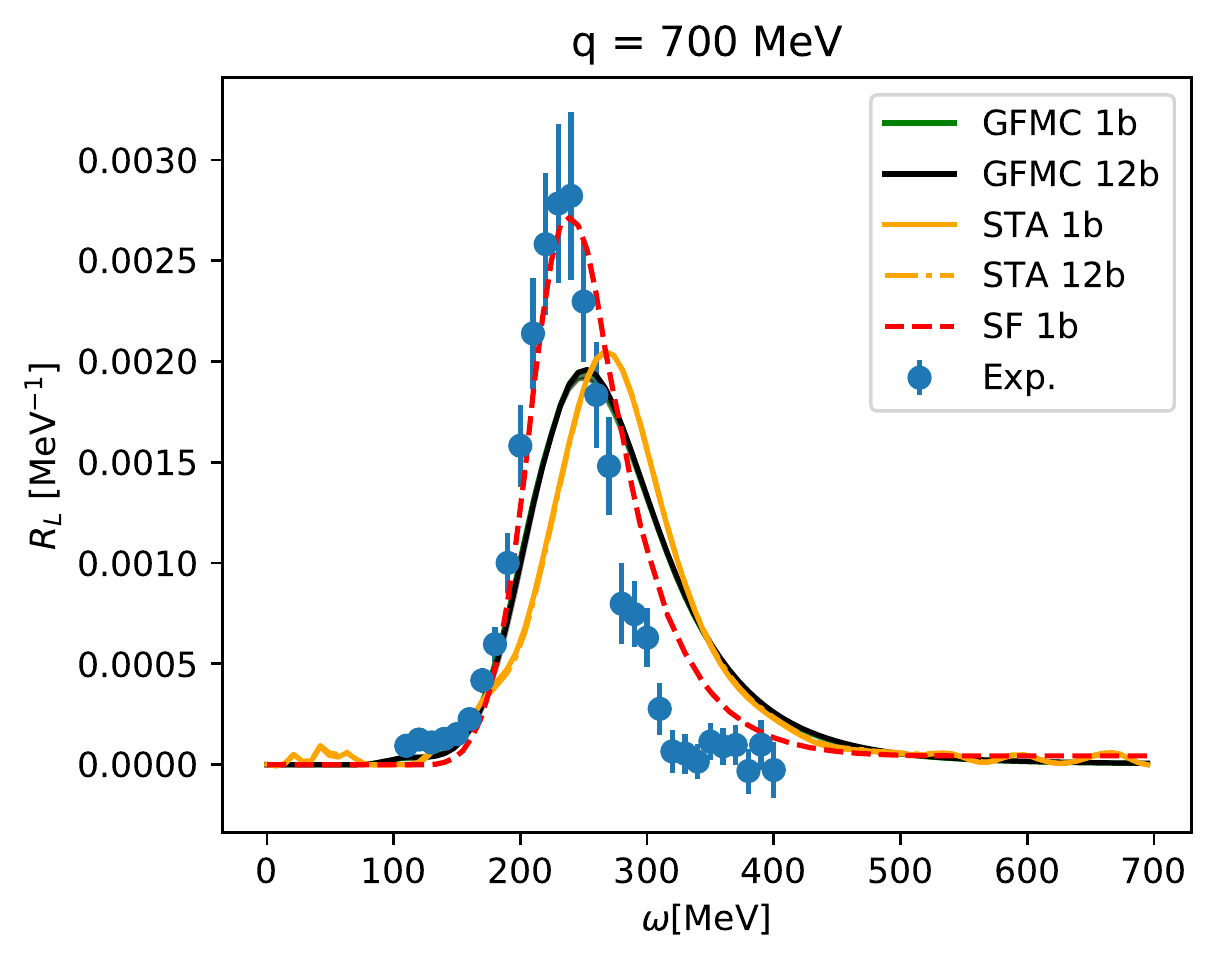} 
    \end{subfigure}
    \hspace{0.1cm}
    \begin{subfigure}[t]{0.48\textwidth}
    \includegraphics[width=\linewidth]{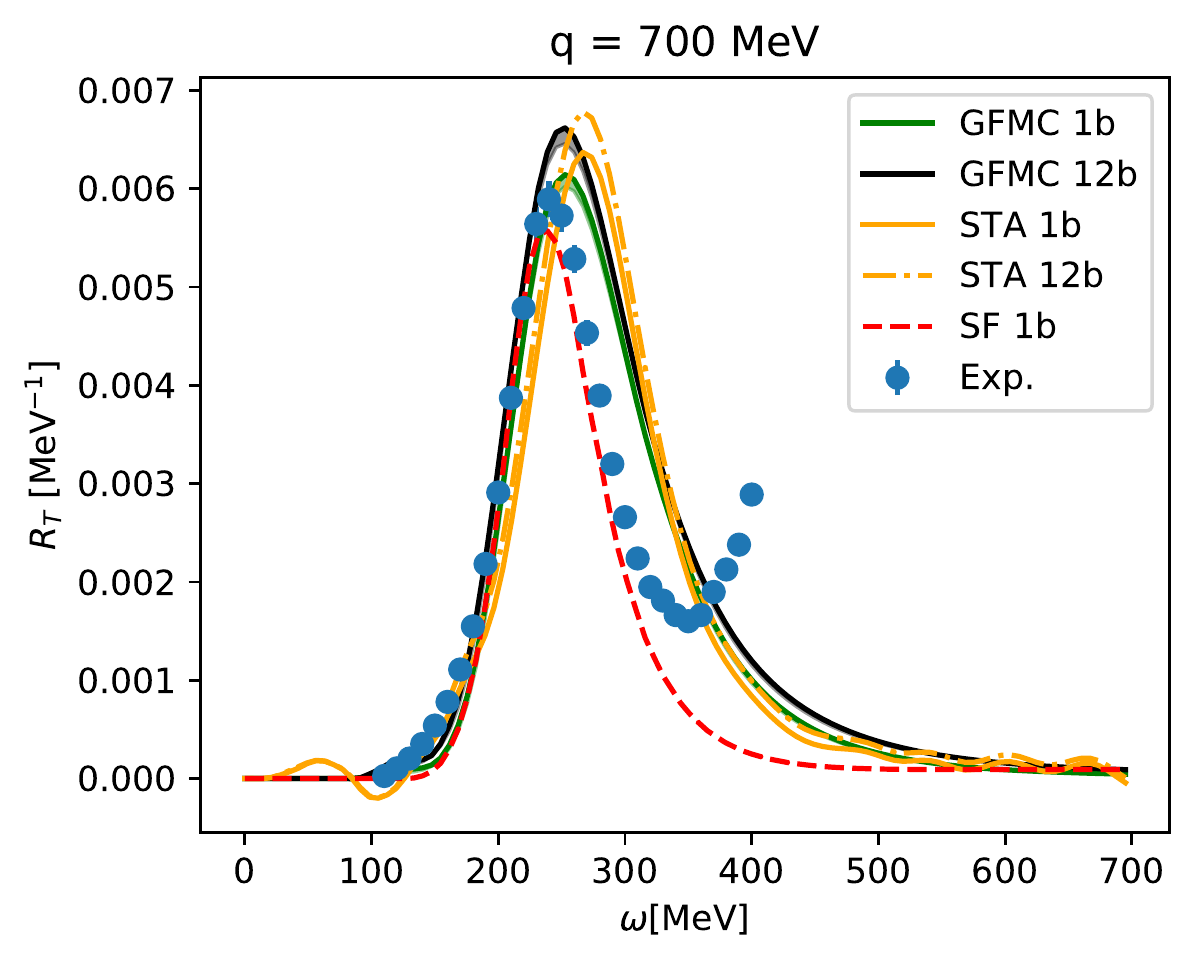} 
    \end{subfigure}
\caption{Longitudinal and transverse response functions of $^3$He. }
\label{fig:resposponse_he3}
\end{figure*}

The different panels of Fig.~\ref{fig:resposponse_he3} display the longitudinal (left) and transverse (right) electromagnetic responses of $^3$He for different values of the momentum transfer $|{\bf q}|$. The blue represent the experimental data of Ref.~\cite{Carlson:2001mp}. The black and green curve correspond to the GFMC one- and one- plus two-body current contributions. The yellow solid and dashed curves display the STA one- and one- plus two-body current calculations and the red dashed line show the SF results, where only the one-body current operator has been included. 
At low momentum transfer, the GFMC results exhibit the correct behavior in both the longitudinal and transverse channel, proving to be in excellent agreement with experiments. Once the elastic contributions are subtracted and the correct behavior at threshold has been enforced, the STA calculations are very close to the GFMC ones. Final state interactions, not included in the current SF calculations, are relevant at $q=300$ MeV. Neglecting these corrections yields an excess of strength in the SF results with respect to the experimental data for the longitudinal response and a shift of the quasiealstic peak toward too large $\omega$ values. Note that, in this work we applied a quenching factor to the SF response functions by subtracting the incoherent contribution of the elastic form factor from the sum rule, corresponding to Eq.~\eqref{eq:eucl_elastic} at $\tau=0$. This effect is more significant for the longitudinal response functions at ${\bf q}$ up to 400 MeV and leads to a quenching of the strength of the response.
The elastic contribution is significantly smaller in the transverse channel and for this reason it has not been subtracted from the transverse response functions obtained within the SF approach. Within the STA, the elastic contribution was found to be negligible in the transverse channel, and in the longitudinal for values of the momentum transfer $q \ge 700$ MeV.
In the STA, the correct behavior at threshold has been enforced to correctly reproduce the transverse response functions at $q = 300$ and $q=400$ MeV. In the GFMC and STA calculations, two-body currents provide the enhancement required to correctly reproduce the transverse data. The SF results, based on the single nucleon current alone, slightly underestimate the data.

The discrepancies between data and the SF results are also visible for $q=500$ MeV because of the missing final-state interaction (FSI) corrections, while for this kinematics the GFMC and STA results nicely agree among each other and with the experiment. 
At $q=700$ MeV, relativistic corrections to both the kinematics and the current operators become sizable and considerably narrow the width of the quasielastic peak. The GFMC and the STA include relativistic corrections in the current operator up to $\mathcal{O}(q^2/m^2)$. However, this is not sufficient to correctly reproduce the position, the height, and the width of the peak in neither the longitudinal nor the transverse channel. On the other hand, for large momentum transfer the factorization of the hadronic vertex is expected to be a good approximation and the SF results with relativistic kinematics correctly reproduce experimental data in both panels.

\begin{figure*}[h!]
    \begin{subfigure}[t]{0.48\textwidth}
    \includegraphics[width=\linewidth]{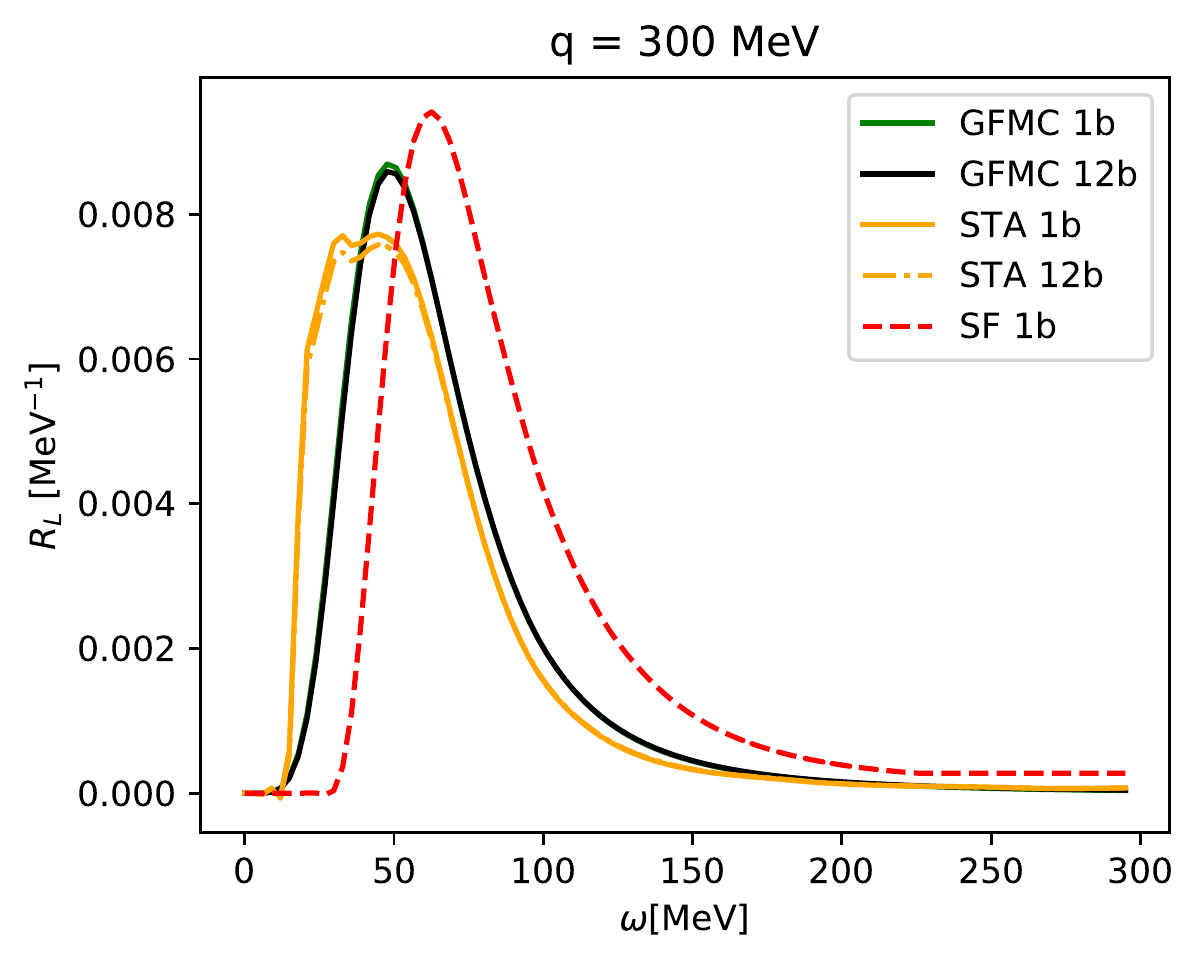} 
    \end{subfigure}
    \hspace{0.1cm}
    \begin{subfigure}[t]{0.48\textwidth}
    \includegraphics[width=\linewidth]{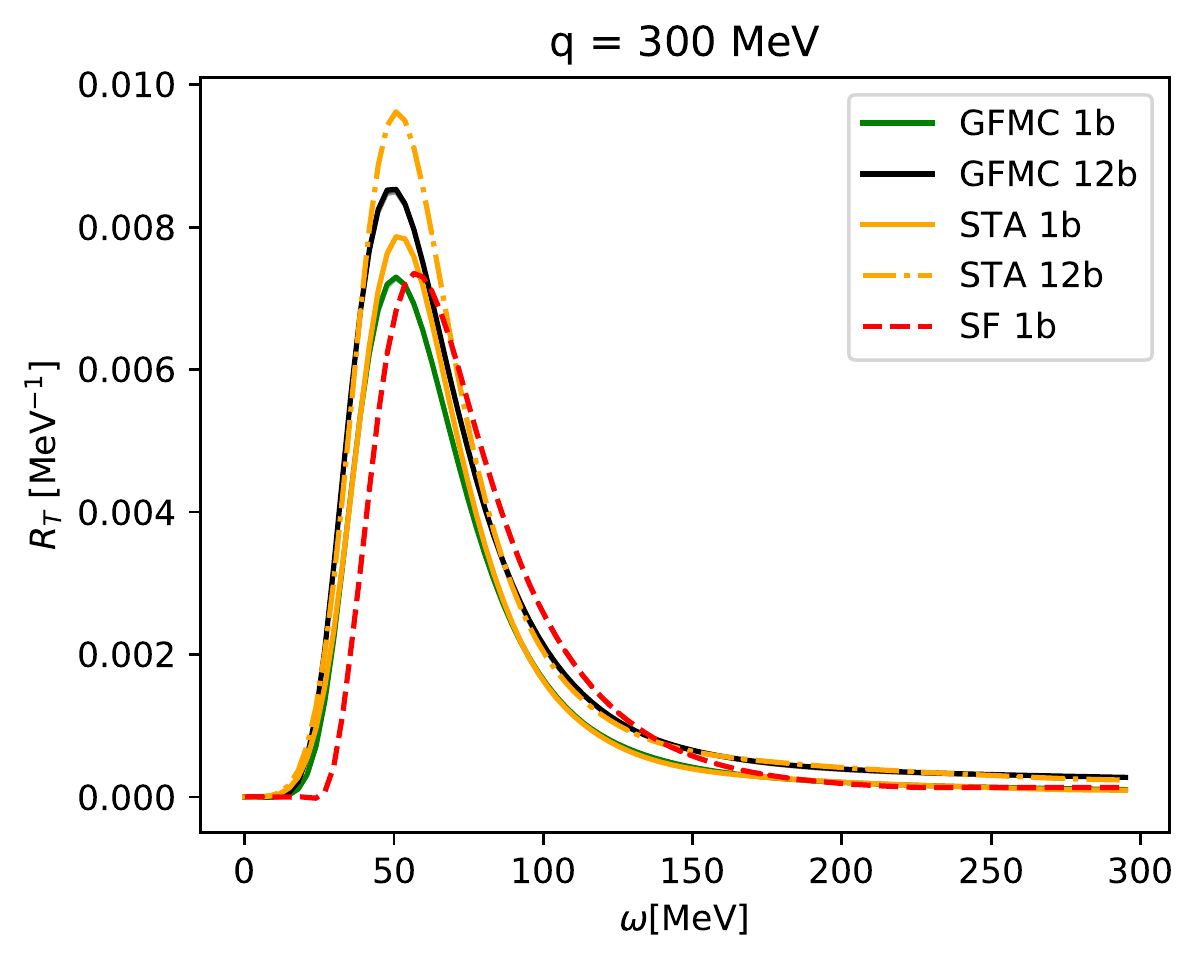} 
    \end{subfigure}
    \begin{subfigure}[t]{0.48\textwidth}
    \includegraphics[width=\linewidth]{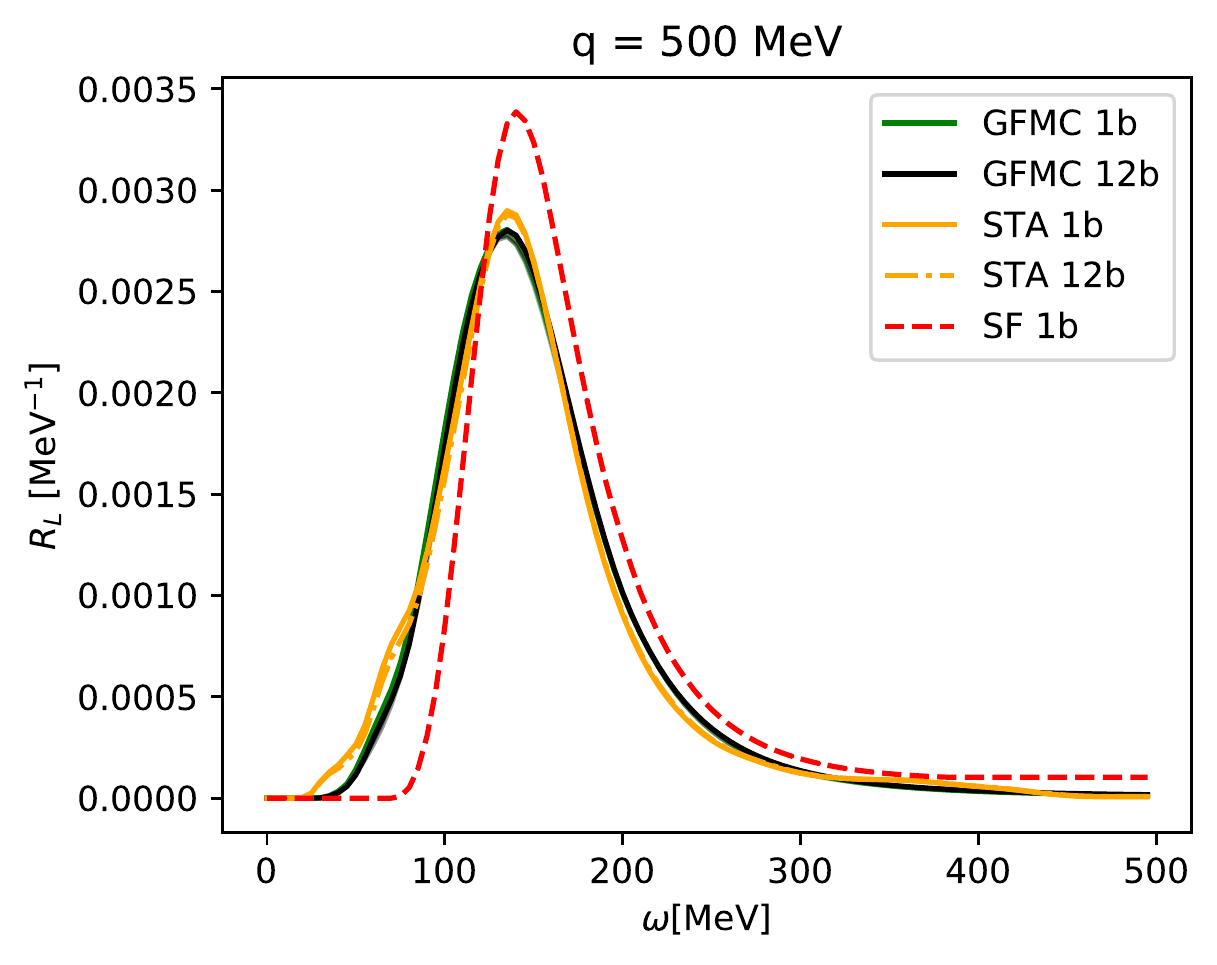} 
    \end{subfigure}
    \hspace{0.1cm}
    \begin{subfigure}[t]{0.48\textwidth}
    \includegraphics[width=\linewidth]{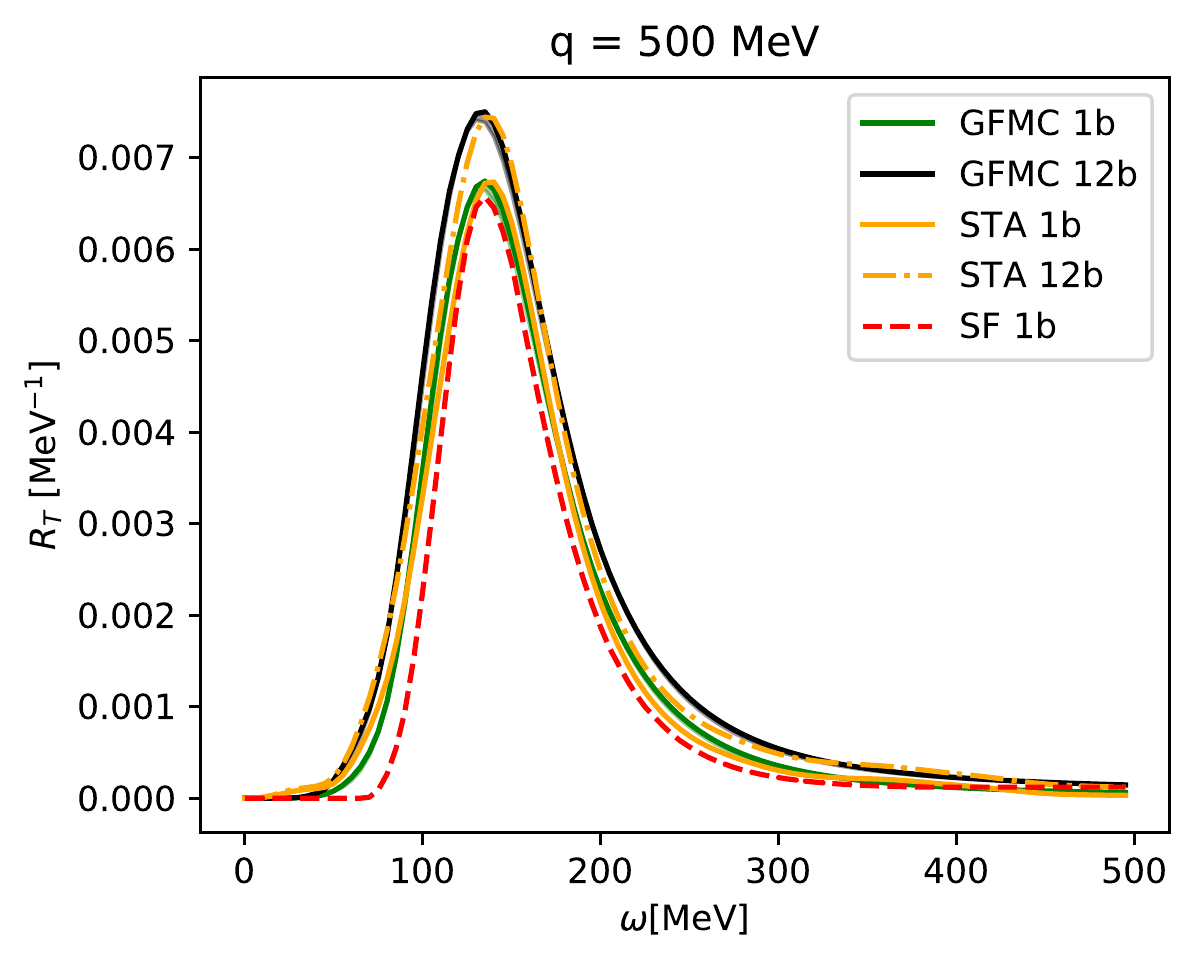} 
    \end{subfigure}
    \begin{subfigure}[t]{0.48\textwidth}
    \includegraphics[width=\linewidth]{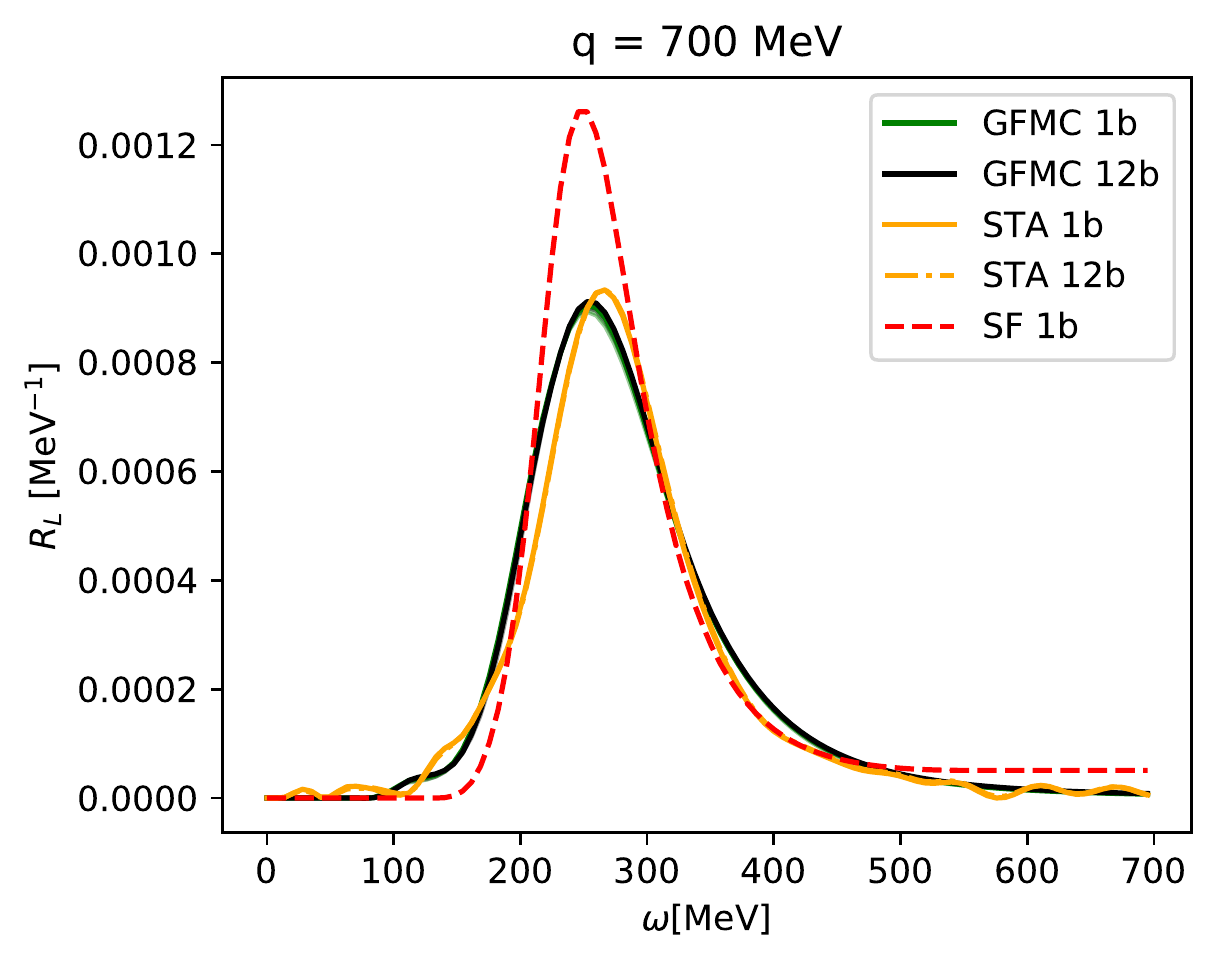} 
    \end{subfigure}
    \hspace{0.1cm}
    \begin{subfigure}[t]{0.48\textwidth}
    \includegraphics[width=\linewidth]{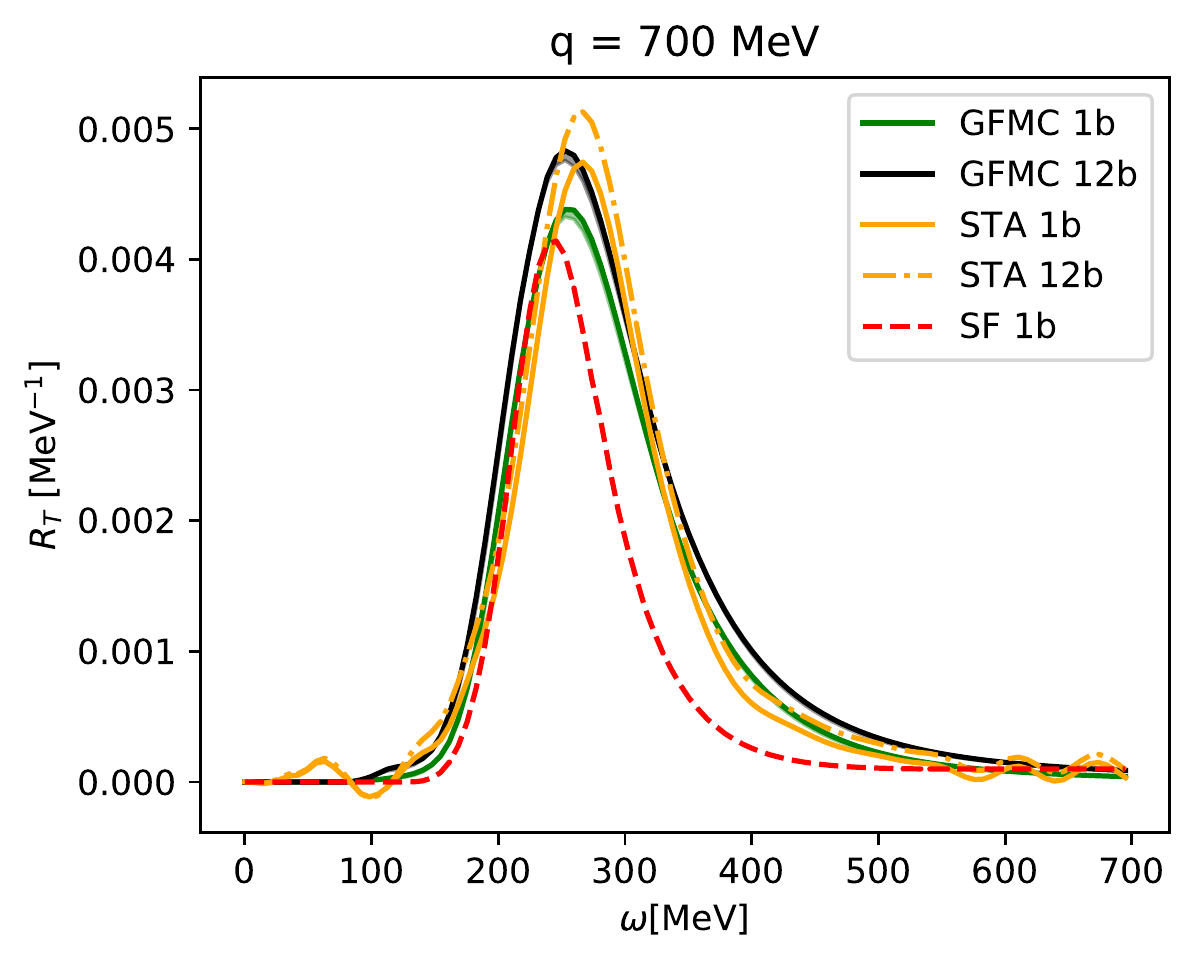} 
    \end{subfigure}
\caption{Longitudinal and transverse response functions of $^3$H. }
\label{fig:resposponse_h3}
\end{figure*}

In Fig.~\ref{fig:resposponse_h3}, we present the longitudinal and transverse responses of $^3$H, for the same values of the momentum transfer as in Fig.~\ref{fig:resposponse_he3}. To the best of our knowledge, there are no experimental data available for the $^3$H electromagnetic response functions at these values of momentum transfer. The general behavior of the theoretical curves is analogous to what observed for $^3$He. However, even after subtracting the spurious elastic contribution from the STA and SF results, the three different approaches exhibit discrepancies in the position and strength of the peaks for $q=$ 300 MeV. The absence of FSI corrections leads to a shift toward large $\omega$ in the SF results for both $q=300$ and $q=500$ MeV. The STA and GFMC results nicely agree for $q=500$ and $q=700$ MeV. Analogously with the $^3$He responses, for the latter values of the momentum transfer, relativistic corrections become relevant, as apparent when comparing the essentially nonrelativistic GFMC and STA results with those obtained within the SF approach. 

\begin{figure*}[h!]
    \begin{subfigure}[t]{0.48\textwidth}
    \includegraphics[width=\linewidth]{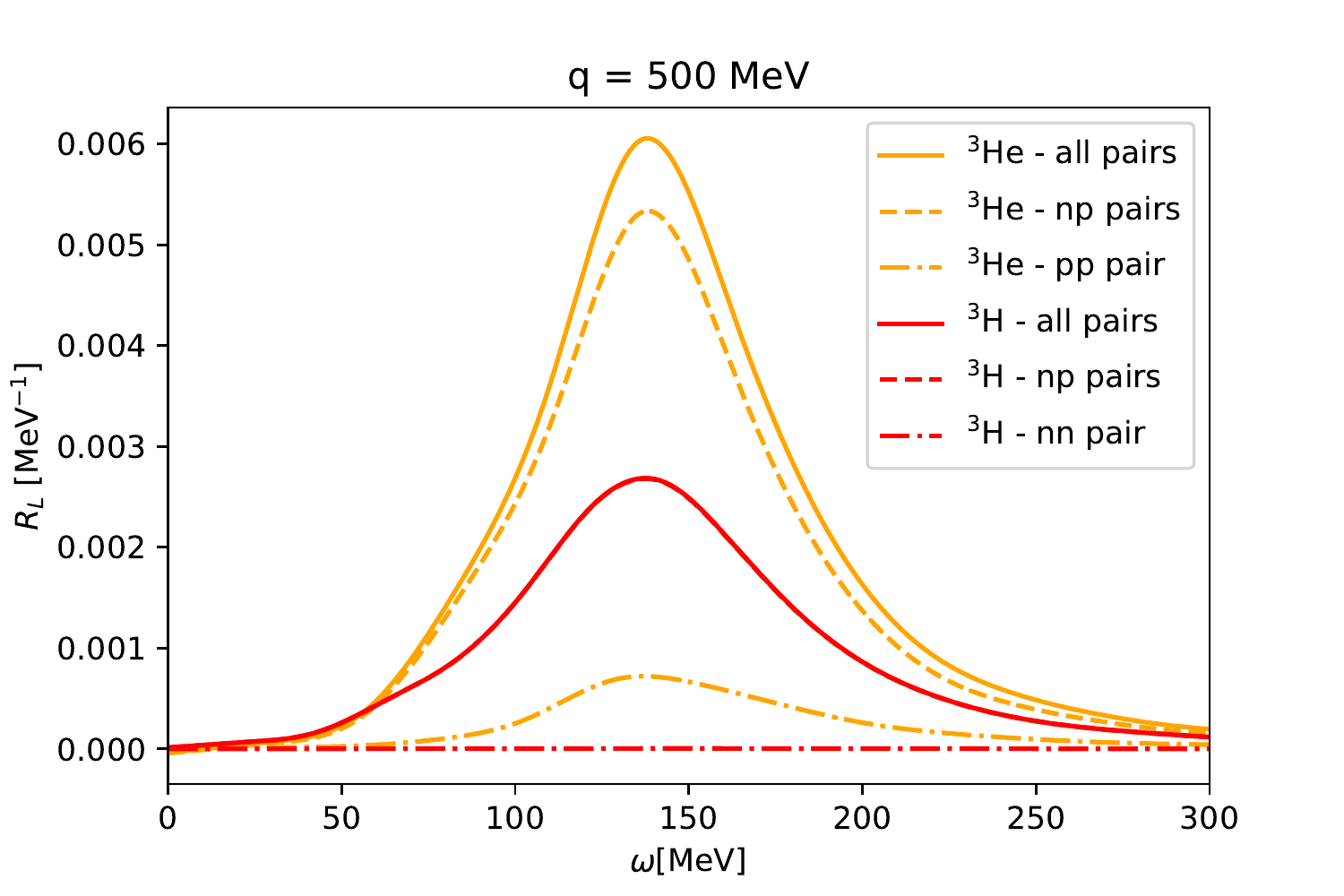} 
    \end{subfigure}
    \hspace{0.1cm}
    \begin{subfigure}[t]{0.48\textwidth}
    \includegraphics[width=\linewidth]{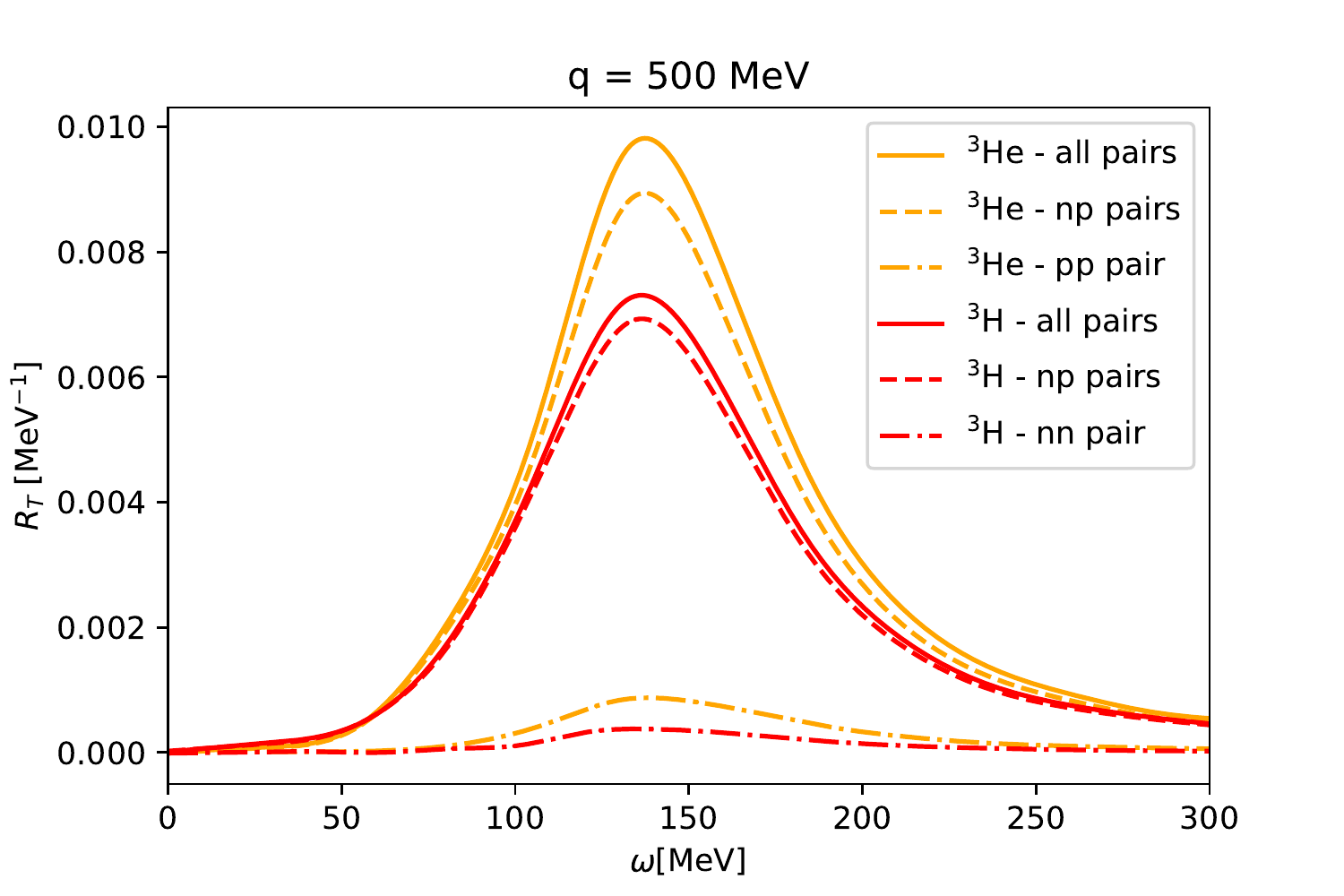} 
    \end{subfigure}
    \caption{STA Longitudinal and transverse response functions of $^3$H and $^3$He and contribution of different pairs. Solid lines are full responses, dashed lines are np pairs, dashed-dotted are nn and pp pairs respectively.}
\label{fig:resposponse_pairs}
\end{figure*}

In Fig.~\ref{fig:resposponse_pairs} we present the contributions of the different nucleon-nucleon pairs for $^3$H and $^3$He response functions, compared to the total responses. 
The data has been smoothed to remove the numerical artifacts that can be seen in the responses obtained in the STA for low and high values of $\omega$. For all practical purposes the same procedure can be applied to the data in Figs.~\ref{fig:resposponse_he3} and \ref{fig:resposponse_h3}.
For both the longitudinal and transverse responses the dominant contribution comes from $np$ pairs. The same is true for the effect of two-nucleon currents.
The longitudinal response (left panel) in $^3$He has a small contribution given only by the  $pp$ pair, while the contribution of the $nn$ pair in $^3$H is negligible, since the charge form factor of the neutron is very small.
In the transverse case (right panel), both $pp$ and $nn$ pairs give a small contribution to the total responses.

\begin{figure*}[h!]
    \begin{subfigure}[t]{0.48\textwidth}
    \includegraphics[width=\linewidth]{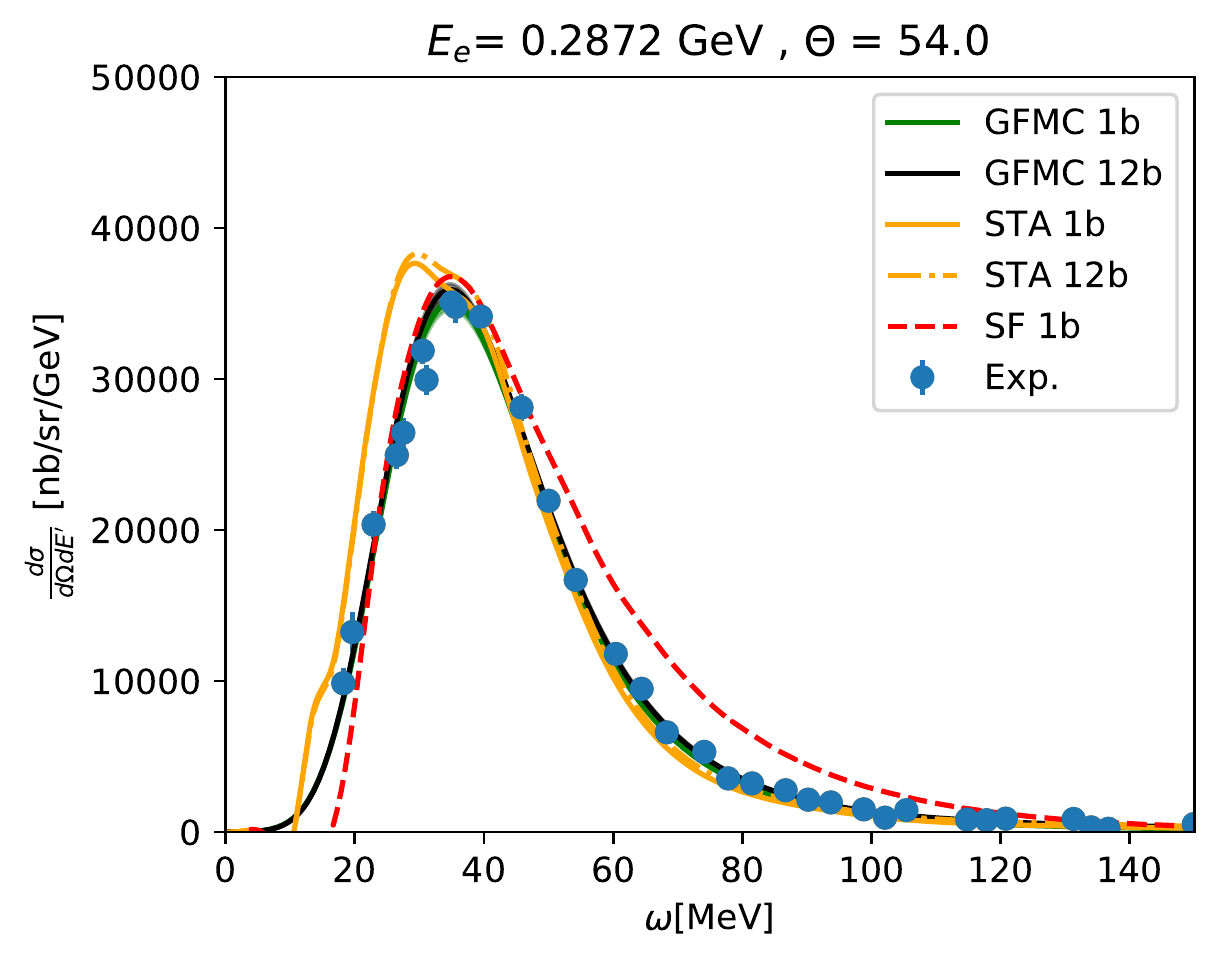} 
    \end{subfigure}
    \hspace{0.1cm}
    \begin{subfigure}[t]{0.48\textwidth}
    \includegraphics[width=\linewidth]{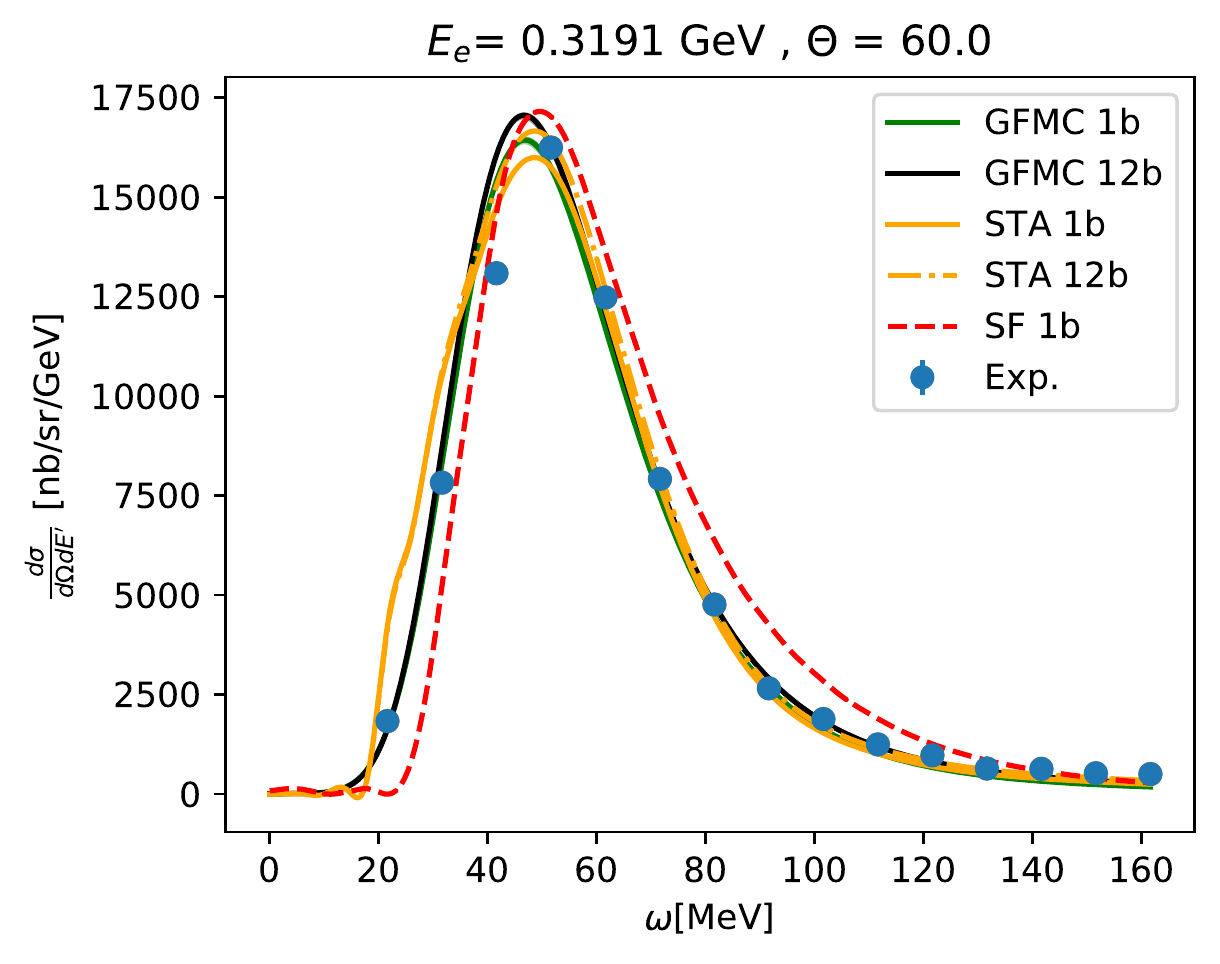} 
    \end{subfigure}
    \begin{subfigure}[t]{0.48\textwidth}
    \includegraphics[width=\linewidth]{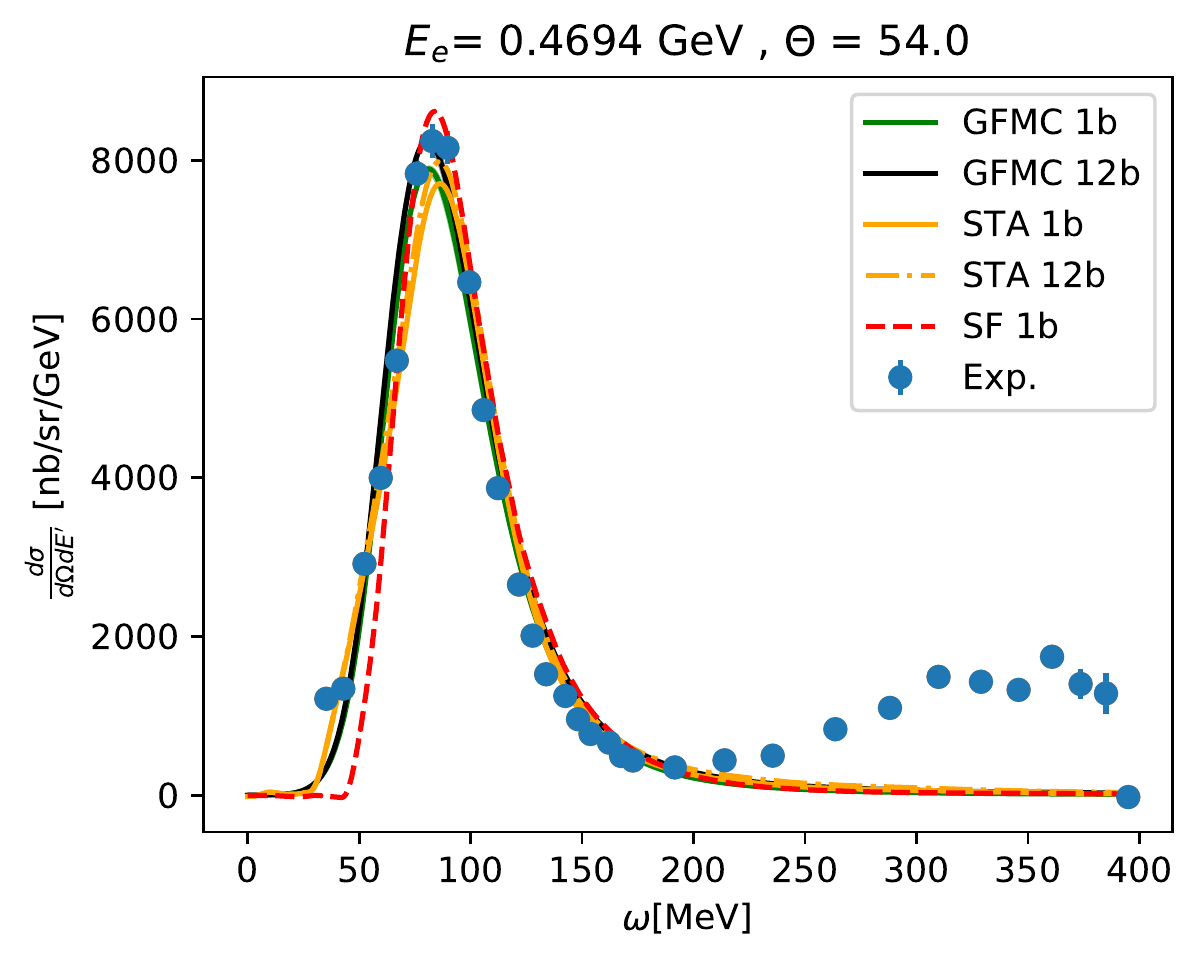} 
    \end{subfigure}
    \hspace{0.1cm}
    \begin{subfigure}[t]{0.48\textwidth}
    \includegraphics[width=\linewidth]{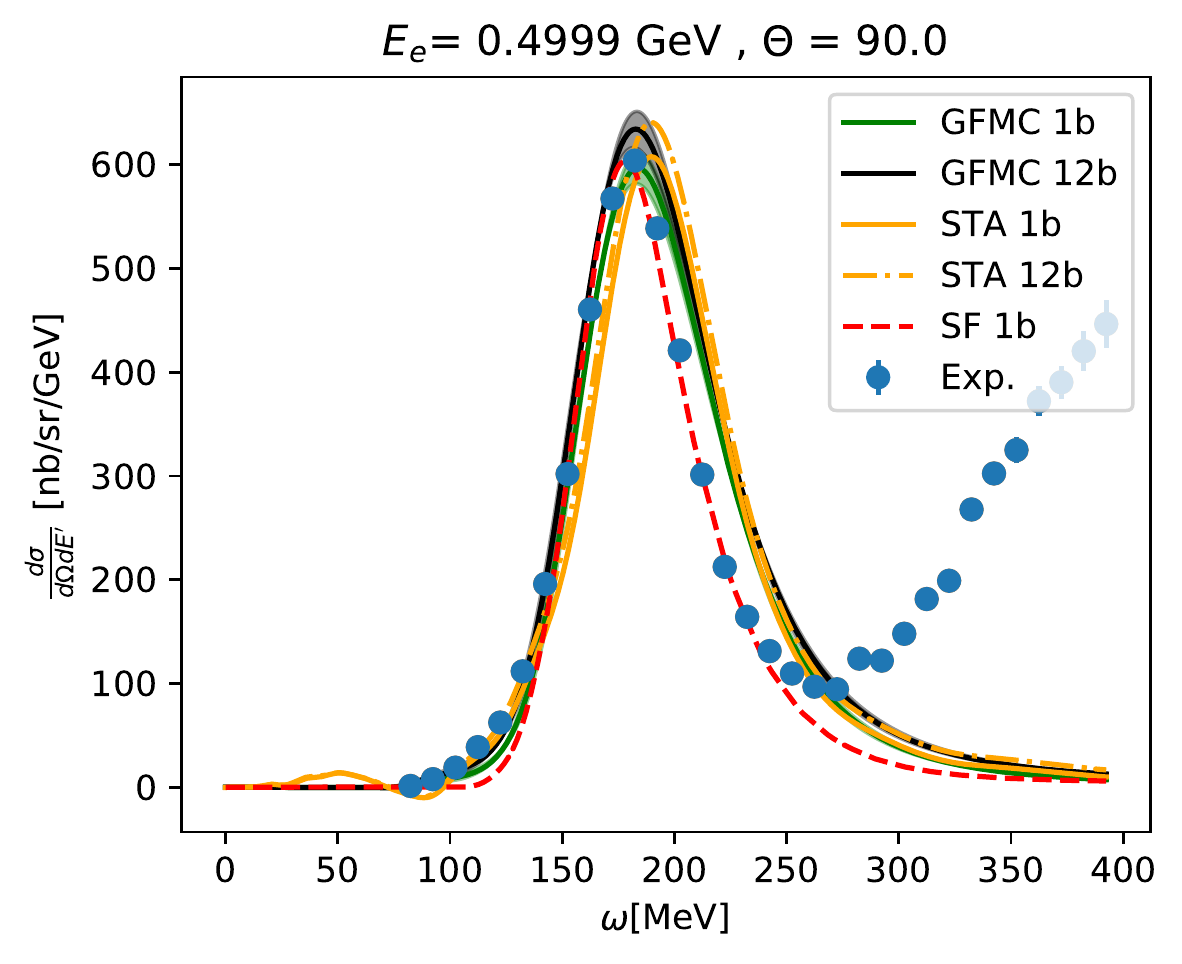} 
    \end{subfigure}
    \begin{subfigure}[t]{0.48\textwidth}
    \includegraphics[width=\linewidth]{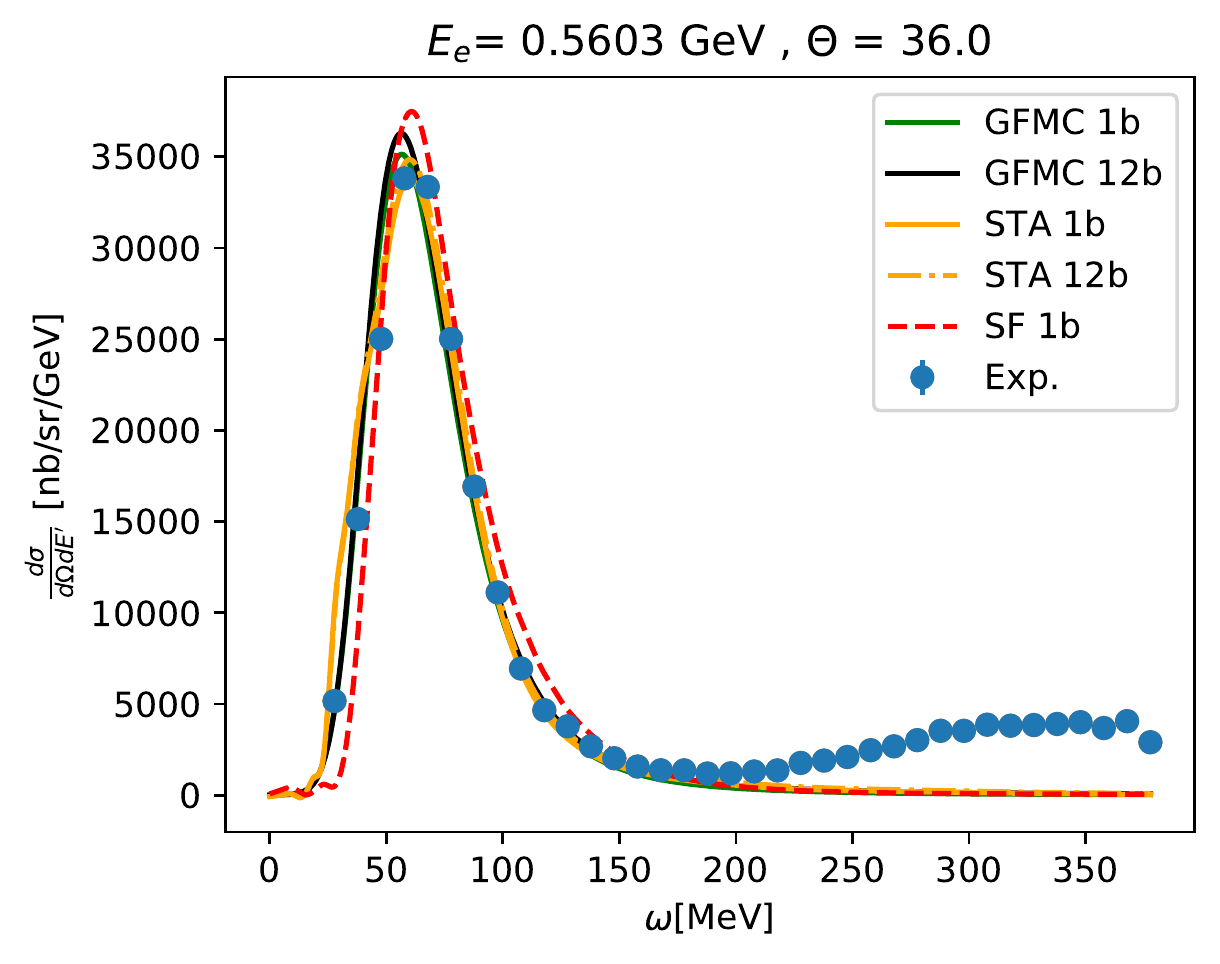} 
    \end{subfigure}
    \hspace{0.1cm}
    \begin{subfigure}[t]{0.48\textwidth}
    \includegraphics[width=\linewidth]{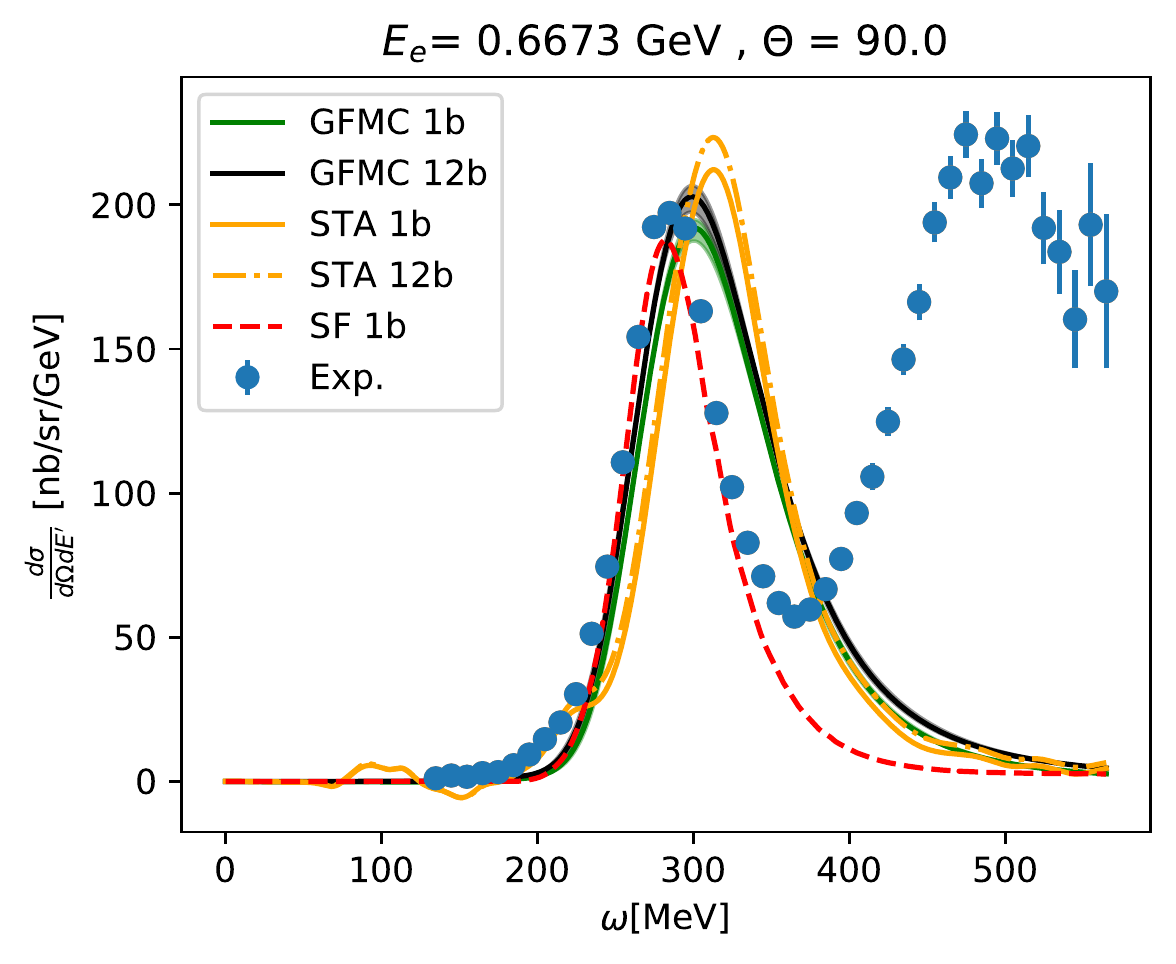} 
    \end{subfigure}
\caption{Inclusive double-differential cross sections for electron scattering on $^3$He. }
\label{fig:cross_sec_he3}
\end{figure*}

\begin{figure*}[h!]
    \begin{subfigure}[t]{0.48\textwidth}
    \includegraphics[width=\linewidth]{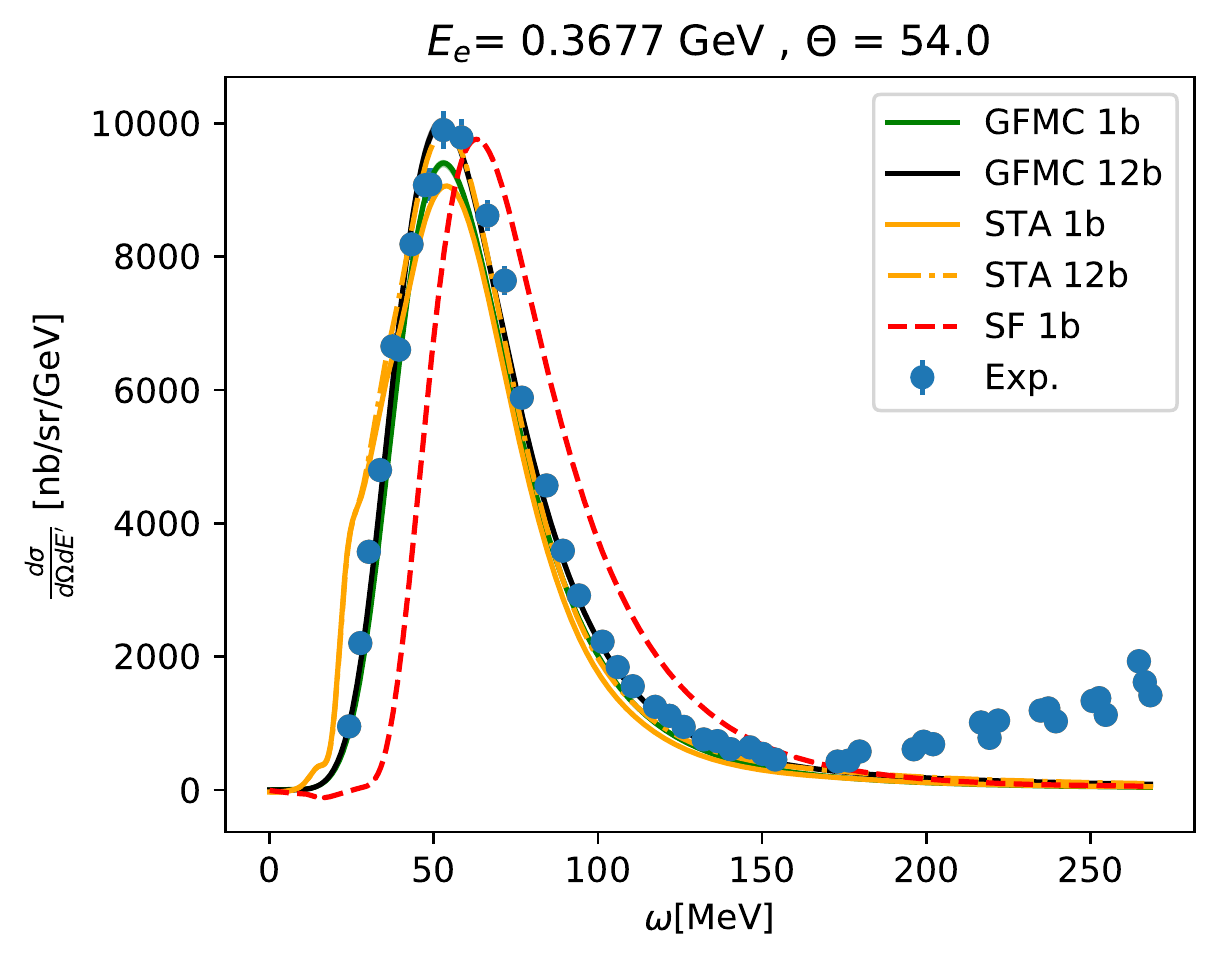} 
    \end{subfigure}
    \hspace{0.1cm}
    \begin{subfigure}[t]{0.48\textwidth}
    \includegraphics[width=\linewidth]{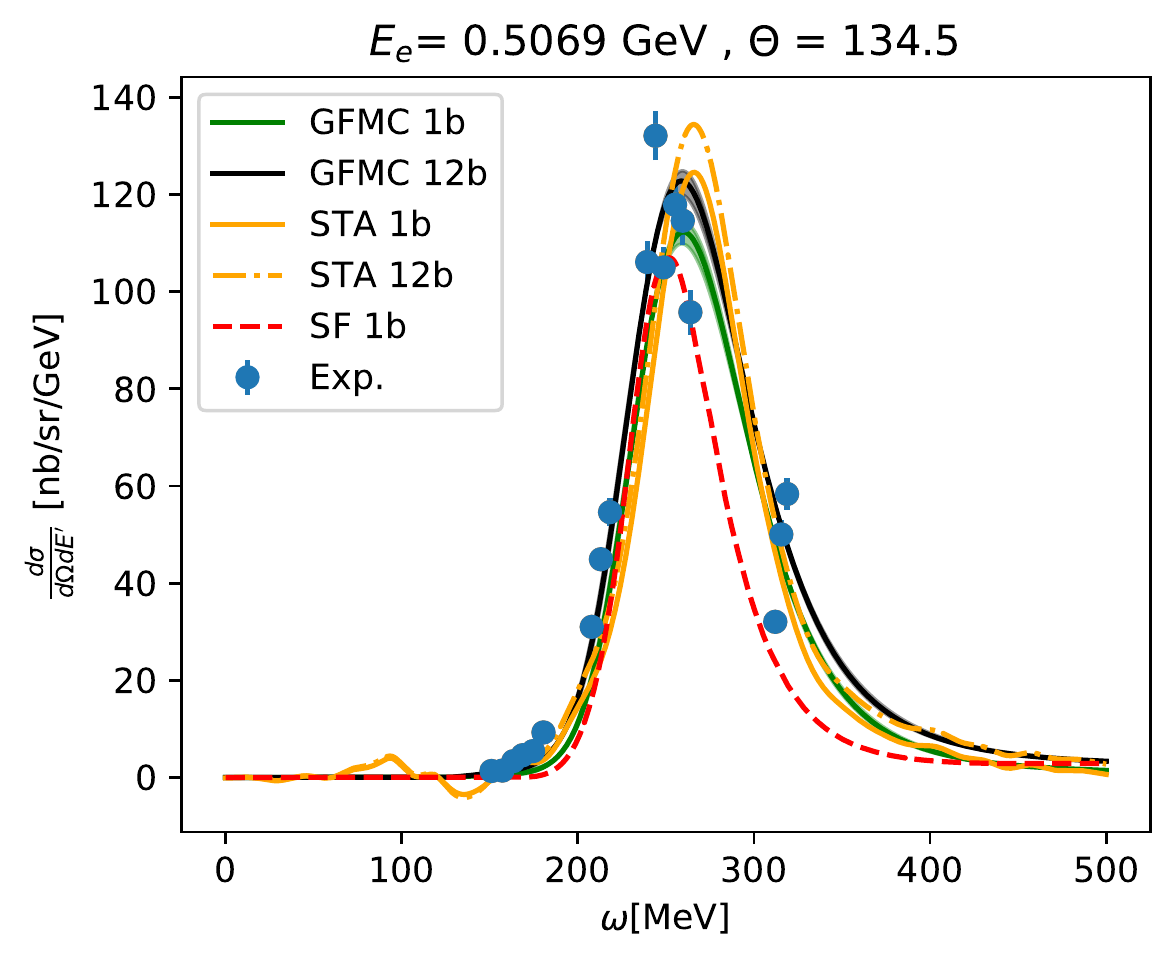} 
    \end{subfigure}
    \begin{subfigure}[t]{0.48\textwidth}
    \includegraphics[width=\linewidth]{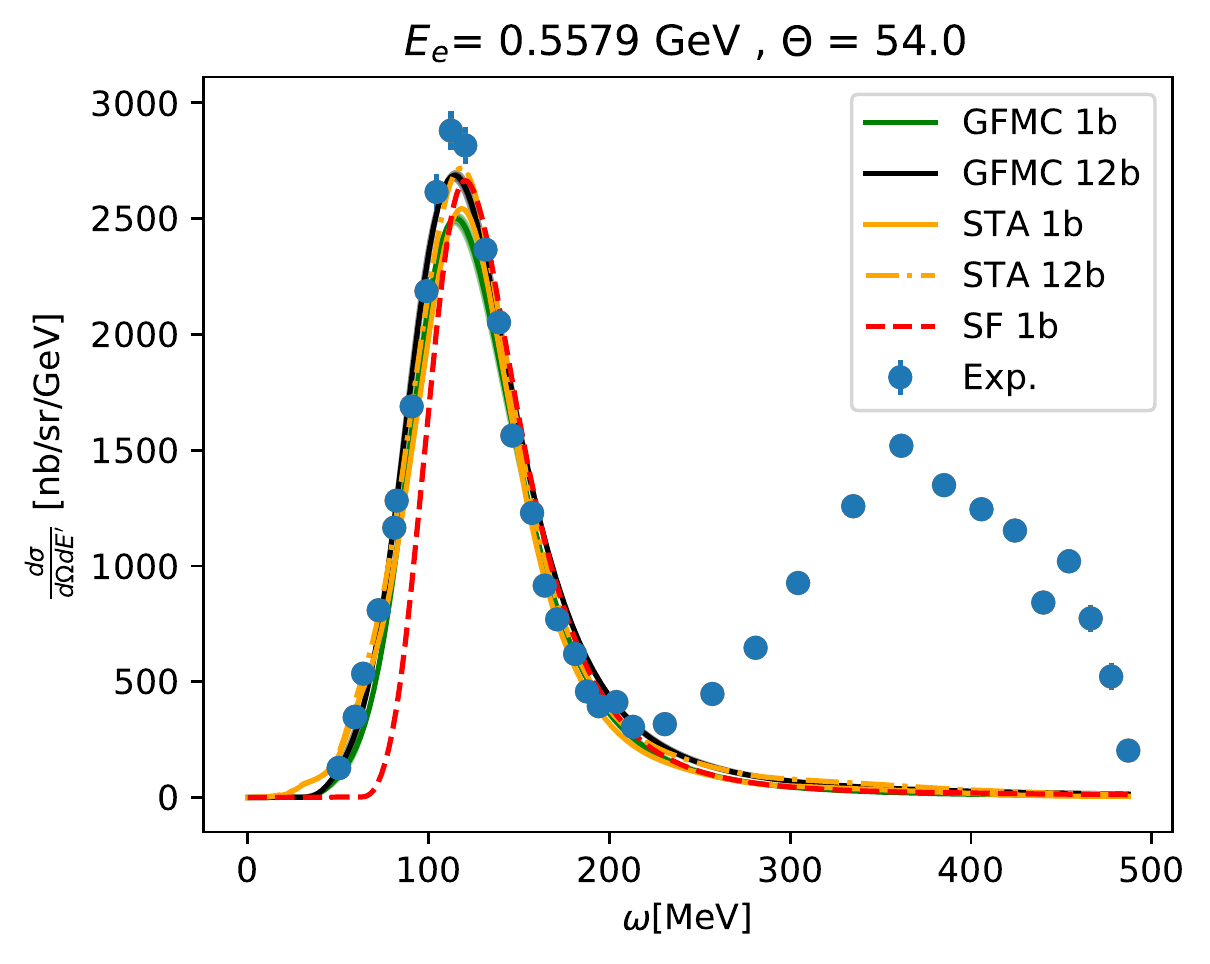} 
    \end{subfigure}
    \hspace{0.1cm}
    \begin{subfigure}[t]{0.48\textwidth}
    \includegraphics[width=\linewidth]{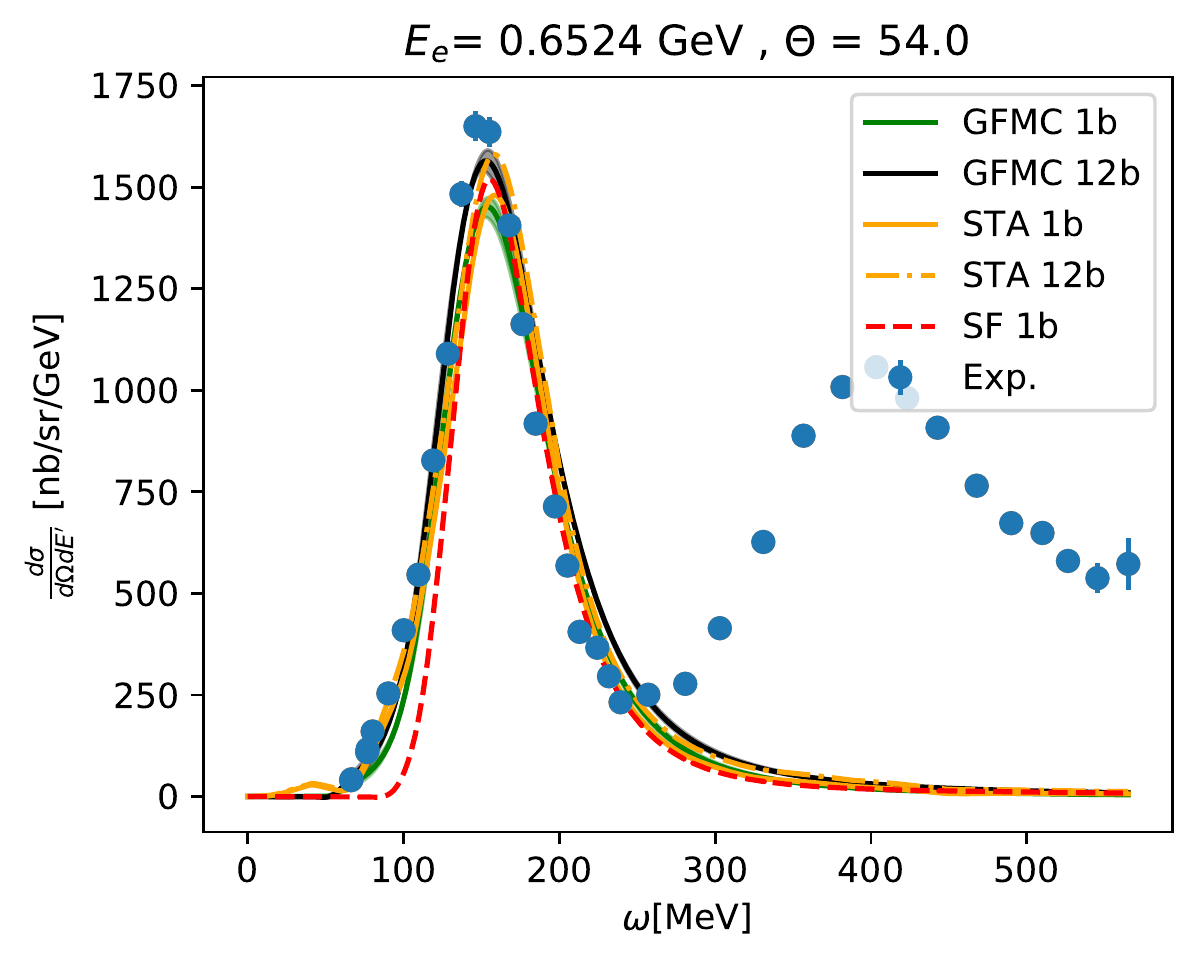} 
    \end{subfigure}
    \begin{subfigure}[t]{0.48\textwidth}
    \includegraphics[width=\linewidth]{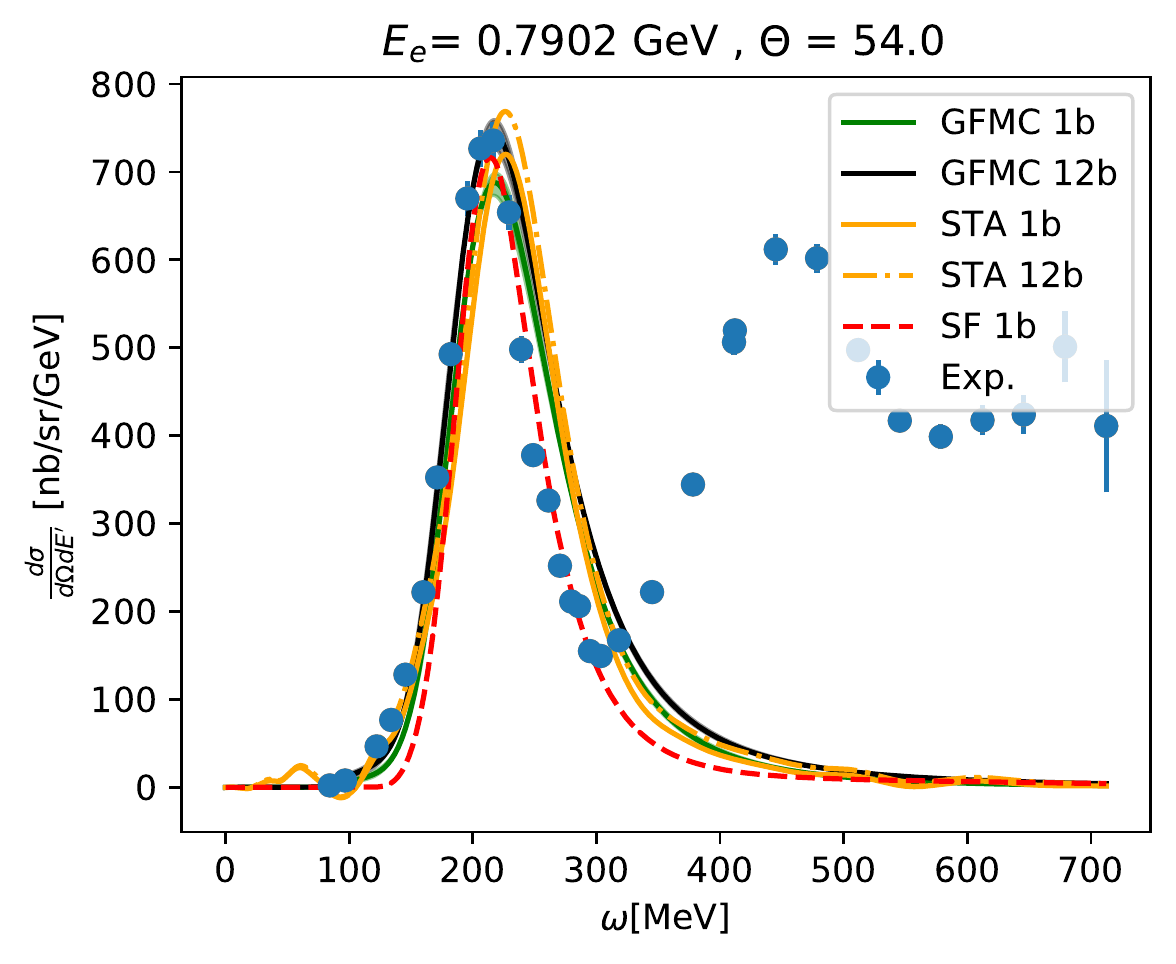} 
    \end{subfigure}
    \hspace{0.1cm}
    \begin{subfigure}[t]{0.48\textwidth}
    \includegraphics[width=\linewidth]{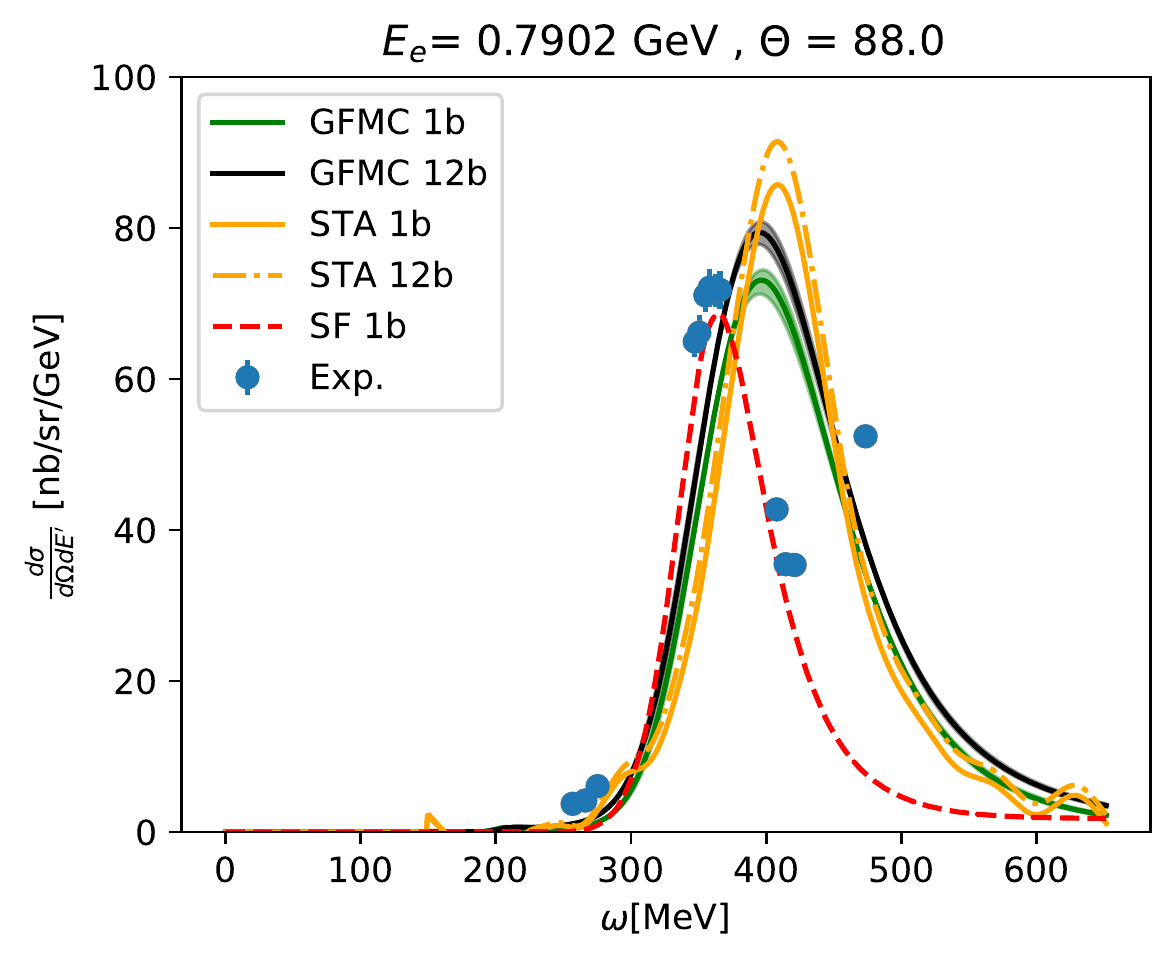} 
    \end{subfigure}
\caption{Inclusive double-differential cross sections for electron scattering on $^3$H. }
\label{fig:cross_sec_h3}
\end{figure*}

The inclusive electron scattering cross section on $^3$He and $^3$H targets are shown in Figs.~\ref{fig:cross_sec_he3} and \ref{fig:cross_sec_h3}, respectively. The various plots correspond to different values of the beam energy and scattering angles and display the double-differential inclusive cross sections as a function of the energy transfer. The curves and colors scheme adopted to represent experimental data, GFMC, STA, and SF theoretical results are the same as in Fig.~\ref{fig:resposponse_he3}. 
The comparison between experiment and theory is restricted to the quasielastic-peak region; reproducing the experimental data at larger $\omega$ would require dealing with pion-production mechanisms. The inclusion of explicit-pion degrees of freedom in the GFMC method is in its infancy~\cite{Madeira:2018ykd} and computing the relevant Euclidean response functions involves nontrivial challenges for the future. On the other hand, since the STA involves only two active nucleons in the final state, it is amenable to the inclusion of pion production channels. As with the SF approach, both the pion-production mechanism and fully relativistic two-body currents have been included --- see Ref.~\cite{Rocco:2019gfb} for recent $^{12}$C calculations. However, more work is required to deal with isospin-asymmetric nuclei, such as $^3$H and $^3$He.  

Neglecting the binding in the initial state and FSI, at the quasielastic peak, the energy and the momentum transfer are related by $\omega_{\rm QE}\simeq ({\bf q}^2_{\rm QE}+m_N^2)^{1/2} - m_N$. For the kinematics in which ${\bf q}_{\rm QE}$ is small, we observe a very good agreement between the GFMC and the data, while the STA and SF results show some minor discrepancies with the experiment. For moderate values of the momentum transfer, there is a very good agreement between the three approaches and the data. Note that for large values of the scattering angle, two-body currents  enhance the GFMC and STA results, primarily via the interference with one-body currents, bringing theoretical calculations closer to experimental data. For the $A=3$ nuclei that we are considering in this work, the two-body current contribution amounts to $\lesssim 10 \%$ of the total strength. This effect is known to be more sizable for larger and more compact nuclei, such as $^4$He and $^{12}$C, as discussed in Refs.~\cite{Lovato:2020kba,Lovato:2016gkq, Pastore:2019urn}. 
In the kinematics with large ${\bf q}_{\rm QE}$, where relativistic effects are dominant, the SF calculations are in good agreement with the experiment in the quasielastic region, whereas the GFMC and STA fail to reproduce the position and the width of the peak. 

\section{Conclusions}
\label{sec:conclusions}
We carried out theoretical calculations of the electromagnetic response functions and inclusive cross sections on $^3$H and $^3$He nuclei for a variety of kinematical setups. These observables are relevant for the electron-scattering program conducted at Jefferson Lab and other facilities worldwide, which are designed to investigate short-range aspects of nuclear structure. Our analysis is based on quantum Monte Carlo methods, as they are ideally suited for accurately treating both the long- and short-range components of the nuclear wave function that emerge from realistic two- and three-body interactions. 

We thoroughly benchmarked three approaches: the Green's function Monte Carlo, the short-time approximation and the spectral function. 
The GFMC, which has already been extensively employed to perform {\it virtually exact} calculations of inclusive electron- and neutrino-scattering~\citep{Lovato:2013cua,Rocco:2018tes,Lovato:2020kba} on $^4$He and $^{12}$C, retains the full complexity of nuclear many-body correlations in both the initial and final states of the reaction. The GFMC results for the electromagnetic response functions and cross sections of $^3$H and $^3$He in the quasielastic region are in excellent agreement with experimental data, for low and moderate values of the momentum transfer. In this regard, it is important to include two-body currents, which bring about a $\sim 10\%$ excess strength with respect to the one-body case. At larger values of $\mathbf{q}$, the nonrelativistic GFMC calculations fail to reproduce the correct position and width of the quasielastic peak. A possible way to improve the GFMC results in this kinematical region is to consider relativistic corrections in the kinematics combined with the use of a convenient reference frame, as done in Ref.~\cite{Rocco:2018tes}.

The STA and the SF approaches are based on the factorization of the final hadronic state, which allows one to overcome some of the limitations of the GFMC. The STA fully retains two-nucleon dynamics and accounts for correlations in the initial and final state as well as two-body currents and interference terms. Consistent with the GFMC results, the latter enhance the transverse response functions in the quasielastic region and are needed to reproduce experimental data. Once the correct behavior at threshold is enforced and the spurious elastic contributions are subtracted from the longitudinal responses at values of momentum transfer lower than $\sim 300$ MeV, a good agreement between the STA calculations and experimental data is observed for the electromagnetic responses and cross sections up to moderate values of the momentum transfer. The current version of the STA approach suffers from analogous limitations as the GFMC in the high momentum transfer region that could be remedied using the strategy of Ref.~\cite{Rocco:2018tes}. The STA is amenable to a more direct inclusion of relativistic effects and meson production mechanisms. In this work, the STA algorithm has been applied to the VMC computational method, it is, however, exportable to other QMC methods suited to study larger nuclear systems, {\it e.g.}, the Auxiliary Field Diffusion Monte Carlo method~\citep{Carlson:2014vla}.

We presented a novel algorithm to obtain the SF of $^3$H and $^3$He combining VMC calculations of the nuclear spectroscopic overlaps with two-body momentum distributions. Since the SF formalism can accommodate fully-relativistic kinematic and currents, its predictions for large values of momentum transfer are in better agreement with experiments than both the GFMC and the STA  In particular, the SF results correctly reproduce the width and the position of the quasielastic peak for both the longitudinal and transverse responses of $^3$He at $q=700$ MeV. However, the SF only retains the incoherent contribution to the nuclear response function and cross section. In addition, final state interaction between the struck nucleon and the spectator system are neglected altogether in the present work. As a consequence, there are some discrepancies between the SF results and experiments at low momentum transfer, which are mitigated once the spurious elastic contribution is subtracted from the theoretical calculations. Two-body currents and pion-production amplitudes have already been implemented in the SF formalism for isospin-asymmetric nuclei~\cite{Rocco:2019gfb}. However, they have not been included in the present $^3$H and $^3$He results because of the nontrivial difficulties involved when dealing with light, isospin-asymmetric nuclei. 

Our analysis has shown that the GFMC, STA and SF results for inclusive electron scattering on both $^3$H- and $^3$He agree reasonably well in all the kinematic regions that we considered. The reason for this agreement has to be found in the consistent description of nuclear correlations in the initial target state --- and in the remnant systems in the case of STA and SF methods. The three methods appear to be remarkably close in the region corresponding to $400$ MeV $\lesssim {\bf q}_{\rm QE} \lesssim$ $600$ MeV. In this regime, the factorization of the final state appears to be a reliable approximation and, concurrently, relativistic effects play a relatively minor role. Besides being relevant for the current experimental program, our study paves the way for precise quantification of the uncertainties inherent to factorization schemes. As a follow up of this work, we intend to carry out a similar analysis for $^{12}$C and other $A\leq 12$ nuclei, on the line of Ref.~\cite{Rocco:2016ejr}. 

Our work has also highlighted a few current limitations of these three methods. Relativistic effects in the interaction vertex and in the kinematics of the reaction play an important role in the high momentum transfer regime. While the SF approach takes them into account, work is ongoing to include them in both the GFMC and STA approaches. On the other hand, some developments are required in the SF formalism to account for two-body currents --- including their interference with one-body terms --- in isospin-asymmetric nuclei.  

\clearpage
\FloatBarrier

\section{Acknowledgements}

This work is supported by the U.S.~Department of Energy under contracts DE-SC0021027 (L.A. and S.P.), DE-AC52-06NA25396 (J.C.), and DE-AC02-06CH11357 (A.L. and R.B.W.), and through the Neutrino Theory Network (A.L. and S.P.) and the FRIB Theory Alliance award DE-SC0013617 (S.P.).
The work of J.C., A.L., and R.B.W. is also supported by the NUclear Computational Low-Energy Initiative (NUCLEI) SciDAC project. A.L. is also supported by the DOE Early Career Research Program award.

N.R. is supported by Fermi Research Alliance, LLC under contract number DE-AC02-07CH11359 with the U.S. DOE, Office of Science, Office of High Energy Physics. The many-body calculations were performed on the parallel computers of the Laboratory Computing Resource Center, Argonne National Laboratory, the computers of the Argonne Leadership Computing Facility via the INCITE grant ``Ab-initio nuclear structure and nuclear reactions'', the 2019/2020 ALCC grant ``Low Energy Neutrino-Nucleus interactions'' for the project NNInteractions, and the 2020/2021 ALCC grant ``Chiral Nuclear Interactions from Nuclei to Nucleonic Matter'' for the project ChiralNuc, and on resources provided by the Los Alamos National Laboratory Institutional Computing Program.

\bibliography{biblio}

\end{document}